%% file: HIN-14-009_temp.tex
\pdfoutput=1

\documentclass[11pt,twoside,a4paper,cmspaper,final,collab]{cms-tdr}
\def\svnVersion{403844}\def\cmsCernNoTag{CERN-EP/2017-009}\def\cmsCernDate{\today}\def\cmsMessage{Published in the European Physical Journal C as \href{http://dx.doi.org/10.1140/epjc/s10052-017-4828-3}{\doi{10.1140/epjc/s10052-017-4828-3.}}}

\begin{document}\cmsNoteHeader{HIN-14-009}

\hyphenation{had-ron-i-za-tion}
\hyphenation{cal-or-i-me-ter}
\hyphenation{de-vices}
\RCS$Revision: 401339 $
\RCS$HeadURL: svn+ssh://svn.cern.ch/reps/tdr2/papers/HIN-14-009/trunk/HIN-14-009.tex $
\RCS$Id: HIN-14-009.tex 401339 2017-04-28 07:08:13Z kyolee $
\newlength\cmsFigWidth
\ifthenelse{\boolean{cms@external}}{\setlength\cmsFigWidth{0.85\columnwidth}}{\setlength\cmsFigWidth{0.4\textwidth}}
\ifthenelse{\boolean{cms@external}}{\providecommand{\cmsLeft}{upper}}{\providecommand{\cmsLeft}{left}}
\ifthenelse{\boolean{cms@external}}{\providecommand{\cmsRight}{lower}}{\providecommand{\cmsRight}{right}}

\newcolumntype{K}[1]{>{\centering\arraybackslash}p{#1}}
\newcommand{\mumu}{\ensuremath{\PGmp\PGmm}\xspace}
\renewcommand{\PgUa}{\ensuremath{\Upsilon\text{(1S)}}\xspace}
\renewcommand{\PgUb}{\ensuremath{\Upsilon\text{(2S)}}\xspace}
\renewcommand{\PgUc}{\ensuremath{\Upsilon\text{(3S)}}\xspace}
\newcommand{\pp}{{\ensuremath{\Pp\Pp}}\xspace}
\newcommand{\PbPb}{\ensuremath{\mathrm{PbPb}}\xspace}
\newcommand{\pPb}{\ensuremath{\Pp\mathrm{Pb}}\xspace}
\newcommand{\Pb}{\ensuremath{\mathrm{Pb}}\xspace}
\newcommand{\sqrts}{\ensuremath{\sqrt{s}}\xspace}
\newcommand{\sqrtsnn}{\ensuremath{\sqrt{\smash[b]{s_{_{\mathrm{NN}}}}}}\xspace}
\newcommand{\rfb}{\ensuremath{R_{\mathrm{FB}}}\xspace}
\newcommand{\rppb}{\ensuremath{R_{\Pp\mathrm{Pb}}}\xspace}
\newcommand{\ethf}{\ensuremath{E_{\mathrm{T}}^{{\mathrm{HF}}\abs{\eta}>4}}\xspace}
\newcommand{\ycm}{\ensuremath{y_{\mathrm{CM}}}\xspace}
\newcommand{\ljpsi}{\ensuremath{\ell_{\JPsi}}\xspace}
\newcommand{\acb}{\ensuremath{\alpha_{\mathrm{CB}}\xspace}\xspace}
\newcommand{\ncb}{\ensuremath{n_{\mathrm{CB}}}\xspace}
\newcommand{\sigN}{\ensuremath{\sigma_{\text{narrow}}}\xspace}
\newcommand{\sigW}{\ensuremath{\sigma_{\text{wide}}}\xspace}

\cmsNoteHeader{HIN-14-009}

\title{Measurement of prompt and nonprompt \JPsi production in \pp and \pPb collisions at $\sqrtsnn=5.02\TeV$}

\date{\today}

\abstract{
This paper reports the measurement of \JPsi meson production in proton-proton (\pp) and proton-lead (\pPb) collisions at a center-of-mass energy per nucleon pair of $5.02\TeV$ by the CMS experiment at the LHC. The data samples used in the analysis correspond to integrated luminosities of 28\pbinv and 35\nbinv for \pp and \pPb collisions, respectively. Prompt and nonprompt \JPsi mesons, the latter produced in the decay of $\PB$ hadrons, are measured in their dimuon decay channels. Differential cross sections are measured in the transverse momentum range of $2<\pt<30\GeVc$, and center-of-mass rapidity ranges of $\abs{y_{\mathrm{CM}}}<2.4$ (\pp) and $-2.87<y_{\mathrm{CM}}<1.93$ (\pPb). The nuclear modification factor, $R_{\pPb}$, is measured as a function of both \pt and $y_{\mathrm{CM}}$. Small modifications to the \JPsi cross sections are observed in \pPb relative to \pp collisions. The ratio of \JPsi production cross sections in \Pp-going and Pb-going directions, $R_{\mathrm{FB}}$, studied as functions of \pt and $y_{\mathrm{CM}}$, shows a significant decrease for increasing transverse energy deposited at large pseudorapidities. These results, which cover a wide kinematic range, provide new insight on the role of cold nuclear matter effects on prompt and nonprompt \JPsi production.
}

\hypersetup{%
 pdfauthor={CMS Collaboration},
 pdftitle={Measurement of prompt and nonprompt J/psi production in pp and pPb collisions at sqrt(s[NN]) = 5.02 TeV},%
 pdfsubject={CMS},%
 pdfkeywords={CMS, physics, dimuons, pp, pPb, quarkonia, charmonia, J/psi, 5.02 TeV}}

\maketitle

\section{Introduction}
\label{sec:introduction}

It was suggested three decades ago that quark-gluon plasma (QGP) formation would suppress the yield of \JPsi mesons in high-energy heavy ion collisions, relative to that in proton-proton (\pp) collisions, as a consequence of Debye screening of the heavy-quark potential at finite temperature~\cite{Matsui:1986dk}. This QGP signature triggered intense research activity, both experimental and theoretical, on the topic of heavy quarkonium production in nuclear collisions. Experiments at SPS~\cite{Abreu200028,Arnaldi:2007zz}, RHIC~\cite{Adare:2006ns,Adamczyk:2013tvk}, and the CERN LHC~\cite{Khachatryan:2016ypw,Abelev:2013ila} have reported a significant \JPsi suppression in heavy ion collisions compared to the expectation based on \pp data. This suppression is found to be larger for more central collisions over a wide range in rapidity ($y$) and transverse momentum (\pt). In addition, a suppression of different bottomonium states $[\PgUa,\,\PgUb,\,\PgUc]$ has been observed at the LHC in lead-lead (\PbPb) collisions at a center-of-mass energy per nucleon pair of $\sqrtsnn=2.76\TeV$~\cite{Chatrchyan:2012lxa,Khachatryan:2016xxp,Abelev:2014nua}, which appears to be consistent with the suggested picture of quarkonium suppression in the QGP~\cite{Emerick:2011xu,Strickland:2011aa}.

In order to interpret these results unambiguously, it is necessary to constrain the so-called cold nuclear matter effects on quarkonium production, through, \eg, baseline measurements in \pPb collisions. Among these effects, parton distribution functions in nuclei (nPDF) are known to differ from those in a free proton and thus influence the quarkonium yields in nuclear collisions. The expected depletion of nuclear gluon density at small values of the momentum fraction ($x$), an effect known as shadowing, would suppress \JPsi production at forward $y$, corresponding to the \Pp-going direction in \pPb collisions~\cite{Ferreiro:2013pua,Vogt:2015uba}. It has been also suggested that gluon radiation induced by parton multiple scattering in the nucleus can lead to \pt broadening and coherent energy loss, resulting in a significant forward \JPsi suppression in \pPb collisions at all available energies~\cite{Arleo:2012hn,Arleo:2013zua}. These phenomena can be quantified by the nuclear modification factor, \rppb, defined as the ratio of \JPsi cross sections in \pPb collisions over those in \pp collisions scaled by the number of nucleons in the \Pb ion ($\mathrm{A}=208$), and by the \rfb ratio of \JPsi cross sections at forward (\Pp-going direction) over those at backward (\Pb-going direction) rapidities.

In addition to prompt \JPsi mesons, directly produced in the primary interaction or from the decay of heavier charmonium states such as $\Pgy$ and $\chi_\text{c}$, the production of \JPsi mesons includes a nonprompt contribution coming from the later decay of $\PB$ hadrons, whose production rates are also expected to be affected by cold nuclear matter effects~\cite{Kang:2014hha,Sickles:2013yna}. However, neither high-\pt $\PB$ mesons nor b quark jets show clear evidence of their cross sections being modified in \pPb collisions~\cite{Khachatryan:2015uja,Khachatryan:2015sva}. In this respect, the nonprompt component of \JPsi production can shed light on the nature of nuclear effects (if any) on bottom-quark production at low \pt.

At the LHC, \JPsi meson production in \pPb collisions at $\sqrtsnn=5.02\TeV$ has been measured by the ALICE~\cite{Abelev:2013yxa,Adam:2015iga}, ATLAS~\cite{Aad:2015ddl}, and LHCb~\cite{Aaij:2013zxa} collaborations. The \rfb ratio has been determined as functions of rapidity in the center-of-mass frame, \ycm, and \pt. Using an interpolation of the \pp production cross sections at the same collision energy, \rppb has also been estimated in Refs.~\cite{Abelev:2013yxa,Adam:2015iga,Aaij:2013zxa} as functions of \ycm and \pt. A significant suppression of the prompt \JPsi production in \pPb collisions has been observed at forward \ycm and low \pt, while no strong nuclear effects are observed at backward \ycm.

This paper reports an analysis of \JPsi production in \pp and \pPb collisions at $\sqrtsnn=5.02\TeV$, using data collected with the CMS detector in 2013 (\pPb) and in 2015 (\pp). The \JPsi mesons with $2<\pt<30\GeVc$ are measured via their dimuon decay channels in ranges of $\abs{\ycm}<2.4$ in \pp and $-2.87<\ycm<1.93$ in \pPb collisions. The corresponding values of $x$ range from $10^{-4}$, at forward \ycm and low \pt, to $10^{-2}$, at backward \ycm and higher \pt. Both \rppb and \rfb are measured as functions of \ycm and \pt. The latter ratio is also studied as a function of the event activity in \pPb collisions, as characterized by the transverse energy deposited in the CMS detector at large pseudorapidities.

\section{Experimental setup and event selection}
\label{sec:eventselection}

The main feature of the CMS detector is a superconducting solenoid with an internal diameter of 6\unit{m}, providing a magnetic field of 3.8\unit{T}. Within the field volume are the silicon pixel and strip tracker, the crystal electromagnetic calorimeter, and the brass and scintillator hadronic calorimeter. The silicon pixel and strip tracker measures charged particle trajectories in the pseudorapidity range of $\abs{\eta}<2.5$. It consists of 66\unit{M} pixel and 10\unit{M} strip sensor elements. Muons are detected in the range of $\abs{\eta}<2.4$, with detection planes based on three technologies: drift tubes, cathode strip chambers, and resistive plate chambers. The CMS apparatus also has extensive forward calorimetry, including two steel and quartz-fiber Cherenkov hadron forward (HF) calorimeters, which cover $2.9<\abs{\eta}<5.2$. These detectors are used for online event selection and the impact parameter characterization of the events in \pPb collisions, where the term impact parameter refers to the transverse distance between the two centers of the colliding hadrons. A more detailed description of the CMS detector, together with a definition of the coordinate system used and the relevant kinematic variables, can be found in Ref.~\cite{CMS:2008zzk}.

The \pPb data set used in this analysis corresponds to an integrated luminosity of 34.6\nbinv. The beam energies are 4\TeV for \Pp, and 1.58\TeV per nucleon for the \Pb nuclei, resulting in $\sqrtsnn=5.02\TeV$. The direction of the higher-energy \Pp\ beam was initially set up to be clockwise, and was reversed after 20.7\nbinv. As a result of the beam energy difference, the nucleon-nucleon center-of-mass in \pPb collisions is not at rest with respect to the laboratory frame. Massless particles emitted at $\abs{\eta_{\mathrm{CM}}}=0$ in the nucleon-nucleon center-of-mass frame are detected at $\eta_{\text{lab}}=-0.465$ for the first run period (clockwise \Pp\ beam) and $+0.465$ for the second run period (counterclockwise \Pp\ beam) in the laboratory frame; the region $-2.87<\ycm<1.93$ is thus probed by flipping the $\eta$ of one data set so that the \Pp-going direction is always toward positive \ycm. The \pp data set is also collected at the same collision energy with an integrated luminosity of 28.0\pbinv. In this sample, \JPsi mesons are measured over $\abs{\ycm}<2.4$.

In order to remove beam-related background such as beam-gas interactions, inelastic hadronic collisions are selected by requiring a coincidence of at least one of the HF calorimeter towers with more than 3\GeV of total energy on each side of the interaction point. This requirement is not present in \pp collisions which suffer less from photon-induced interactions compared to \pPb collisions. The \pp and \pPb events are further selected to have at least one reconstructed primary vertex composed of two or more associated tracks, excluding the two muons from the \JPsi candidates, within 25\cm from the nominal interaction point along the beam axis and within 2\cm in its transverse plane. To reject beam-scraping events, the fraction of good-quality tracks associated with the primary vertex is required to be larger than 25\% when there are more than 10 tracks per event.

In \pPb collisions, an additional filter~\cite{Chatrchyan:2013nka} is applied to remove events containing multiple interactions per bunch crossing (pileup). After the selection, the residual fraction of pileup events is reduced from 3\% to less than 0.2\%. This pileup rejection results in a 4.1\% signal loss, which is corrected for in the cross section measurements. Since pileup only affects the event activity dependence in \pPb results, no filter is applied in \pp results.

Dimuon events are selected by the level-1 trigger, a hardware-based trigger system requiring two muon candidates in the muon detectors with no explicit limitations in \pt or $y$. In the offline analysis, muons are required to be within the following kinematic regions, which ensure single-muon reconstruction efficiencies above 10\%:
\begin{equation}
\begin{aligned}
&\pt^{\mu}>3.3\GeVc & \text{ for }\abs{\eta_{\text{lab}}^{\mu}}<1.2,\\
&\pt^{\mu}>(4.0-1.1\abs{\eta_{\text{lab}}^{\mu}})\GeVc & \text{ for }1.2\le\abs{\eta_{\text{lab}}^{\mu}}<2.1,\\
&\pt^{\mu}>1.3\GeVc &\text{ for }2.1\le\abs{\eta_{\text{lab}}^{\mu}}<2.4.\\
\end{aligned}
\label{eqn:acccut}
\end{equation}
The muon pairs are further selected to be of opposite charge, to originate from a common vertex with a $\chi^2$ probability greater than 1\%, and to match standard identification criteria~\cite{Chatrchyan:2012xi}.

Simulated events are used to obtain the correction factors for acceptance and efficiency. The Monte Carlo (MC) samples of \JPsi mesons are generated using \PYTHIA 8.209~\cite{Sjostrand:2014zea} for \pp and \PYTHIA 6.424~\cite{Sjostrand:2006za} for \pPb collisions. Generated particles in the \pPb simulation are boosted by $\Delta y=\pm0.465$ to account for the asymmetry of \Pp\ and \Pb beams in the laboratory frame. Samples for prompt and nonprompt \JPsi mesons are independently produced using the D6T~\cite{Bartalini:2010su} and Z2~\cite{Field:2010bc} tunes, respectively. In the absence of experimental information on quarkonium polarization in \pp and \pPb collisions at $\sqrts=5.02\TeV$, it is assumed that prompt \JPsi mesons are produced unpolarized, as observed in \pp collisions at $\sqrts=7\TeV$~\cite{Chatrchyan:2013cla,Aaij:2013nlm,Abelev:2011md}. The nonprompt \JPsi sample includes the polarization ($\lambda_{\theta}\approx-0.4$) determined from a measurement of the exclusive $\PB$ hadron decays ($\PBp, \PBz$, and $\PBzs$) as implemented in \EVTGEN 9.1~\cite{Lange:2001uf}. The \pPb measurements might be affected by physics processes with strong kinematic dependence within an analysis bin, \eg, polarization or energy loss. Such possible physics effects on the final cross sections are not included in the systematic uncertainties, as was done in the previous analyses~\cite{Chatrchyan:2012lxa,Khachatryan:2016xxp}.
The QED final-state radiation from muons is simulated with \PHOTOS 215.5~\cite{Barberio:1993qi}. Finally, the CMS detector response is simulated using \GEANTfour~\cite{Agostinelli:2002hh}.

\section{Analysis procedure}
\label{sec:analysisprocedure}

\subsection{Differential cross section, \texorpdfstring{\rppb}{R\_pPb}, and \texorpdfstring{\rfb}{R\_FB}}
\label{sec:definitionofvariables}

In this paper, three observables analyzed in \JPsi meson decays to muon pairs are reported. First, the cross sections are determined based on
\begin{equation}
\mathcal{B}(\JPsi\to\mumu)\frac{\rd^2\sigma}{\rd\pt\,\rd\ycm} = \frac{N^{\JPsi}_{\text{Fit}}/(\text{Acc}\,\varepsilon)}{\mathcal{L}_{\text{int}}\,\Delta\pt\,\Delta\ycm},
\label{eqn:crosssection}
\end{equation}
where $\mathcal{B}(\JPsi\to\mumu)$ is the branching fraction to the \mumu channel~\cite{pdg:2016}, $N^{\JPsi}_{\text{Fit}}$ is the extracted raw yield of \JPsi mesons in a given $(\pt,\ycm)$ bin, $(\text{Acc}\,\varepsilon)$ represents the dimuon acceptance times efficiency described in Section~\ref{sec:corrections}, and $\mathcal{L}_{\text{int}}$ is the integrated luminosity with the values of $(28.0\pm0.6)$\pbinv for \pp~\cite{CMS-PAS-LUM-16-001} and $(34.6\pm1.2)$\nbinv for \pPb~\cite{CMS-PAS-LUM-13-002} collisions.

The cross sections are measured in up to nine bins in \pt ([2,3], [3,4] [4,5], [5,6.5], [6.5,7.5], [7.5,8.5], [8.5,10], [10,14], [14,30]\GeVc), with the minimum \pt values varying with \ycm ranges as shown in Table~\ref{tab:minptvalue}.

\begin{table}[htbp]
\topcaption{Rapidity intervals and associated minimum \pt values for the \JPsi cross section measurements in \pp and \pPb collisions.}
\label{tab:minptvalue}
\centering
\begin{tabular}{cK{1.6cm}K{1.6cm}}
\hline
\multirow{2}{*}{\ycm} & \multicolumn{2}{c}{Minimum \pt (\GeVcns)} \\
\cline{2-3}
 & \pp & \pPb \\
\hline
 $1.93<\ycm<2.4$ & 2 & N/A \\
 $1.5<\ycm<1.93$ & 4 & 2 \\
 $0.9<\ycm<1.5$ & 6.5 & 4 \\
 $0<\ycm<0.9$ & 6.5 & 6.5 \\
 $-0.9<\ycm<0$ & 6.5 & 6.5 \\
 $-1.5<\ycm<-0.9$ & 6.5 & 6.5 \\
 $-1.93<\ycm<-1.5$ & 4 & 5 \\
 $-2.4<\ycm<-1.93$ & 2 & 4 \\
 $-2.87<\ycm<-2.4$ & N/A & 2 \\
\hline
\end{tabular}
\end{table}

The second observable considered is the nuclear modification factor, calculated as
\begin{equation}
\label{eqn:rpa}
\rppb(\pt,\ycm) = \frac{({\rd^2\sigma}/{\rd\pt\,\rd\ycm})_{\pPb}}{\mathrm{A}({\rd^2\sigma}/{\rd\pt\,\rd\ycm})_{\pp}},
\end{equation}
where $\mathrm{A}=208$ is the number of nucleons in the \Pb nucleus.

The third measurement is the forward-to-backward production ratio for \pPb collisions, defined for positive \ycm by
\begin{equation}
\rfb(\pt,\ycm>0) = \frac{\rd^2\sigma(\pt,\ycm)/\rd\pt\rd\ycm}{\rd^2\sigma(\pt,-\ycm)/\rd\pt\rd\ycm}.
\end{equation}
This variable is a sensitive probe of the dynamics of \JPsi production by comparing nuclear effects in the forward and the backward \ycm hemispheres, since $\rfb(\pt,\ycm)$ is equivalent to $\rppb(\pt,\ycm)/\rppb(\pt,-\ycm)$. In addition, several uncertainties cancel in the \rfb ratio, such as those from the integrated luminosity determination. The minimum \pt values for the \rfb measurement are 5\GeVc for $1.5<\abs{\ycm}<1.93$, and 6.5\GeVc for $\abs{\ycm}<1.5$. The ratio \rfb is also analyzed as a function of \ethf, the transverse energy deposited on both sides of the collisions in the HF calorimeters within the $4<\abs{\eta}<5.2$ range. This energy is related to the impact parameter of the collision. In Table~\ref{tab:hfbin_pa}, the mean value of \ethf and the fraction of events for each bin used in the analysis are computed from minimum bias \pPb events.

\begin{table}[htbp]
\topcaption{Ranges of forward transverse energy, \ethf, their mean values, and associated fractions of \pPb events that fall into each category.}
\label{tab:hfbin_pa}
\centering
\begin{tabular}{ccc }
\hline
 $\ethf\,(\GeVns)$ & $\langle\ethf\rangle$ & Fraction \\
\hline
 0--20 & 9.4 & 73\% \\
 20--30 & 24.3 & 18\% \\
 $>$30 & 37.2 & 9\% \\
\hline
\end{tabular}
\end{table}

\subsection{Signal extraction}
\label{sec:signalextraction}

The signal extraction procedure is similar to that in previous CMS analyses of \pp~\cite{Chatrchyan:2011kc, Khachatryan:2015rra} and \PbPb~\cite{Khachatryan:2016ypw} collisions. The prompt \JPsi mesons are separated from those coming from $\PB$ hadron decays by virtue of the pseudo-proper decay length, $\ljpsi=L_{xy}\,m_{\JPsi}/\pt$, where $L_{xy}$ is the transverse distance between the primary and secondary dimuon vertices in the laboratory frame, $m_{\JPsi}$ is the mass of the \JPsi meson, and \pt is the dimuon transverse momentum. For each \pt, $\ycm$, and event activity bin, the fraction of nonprompt \JPsi mesons (\textit{b fraction}) is evaluated through an extended unbinned maximum likelihood fit to the invariant mass spectrum and \ljpsi distributions of \mumu pairs, sequentially. The invariant mass spectrum is fitted first, and some parameters are initialized and/or fixed. Then, the \ljpsi distribution is fitted.

\begin{figure*}[ht]
\begin{center}
\includegraphics[width=0.48\textwidth]{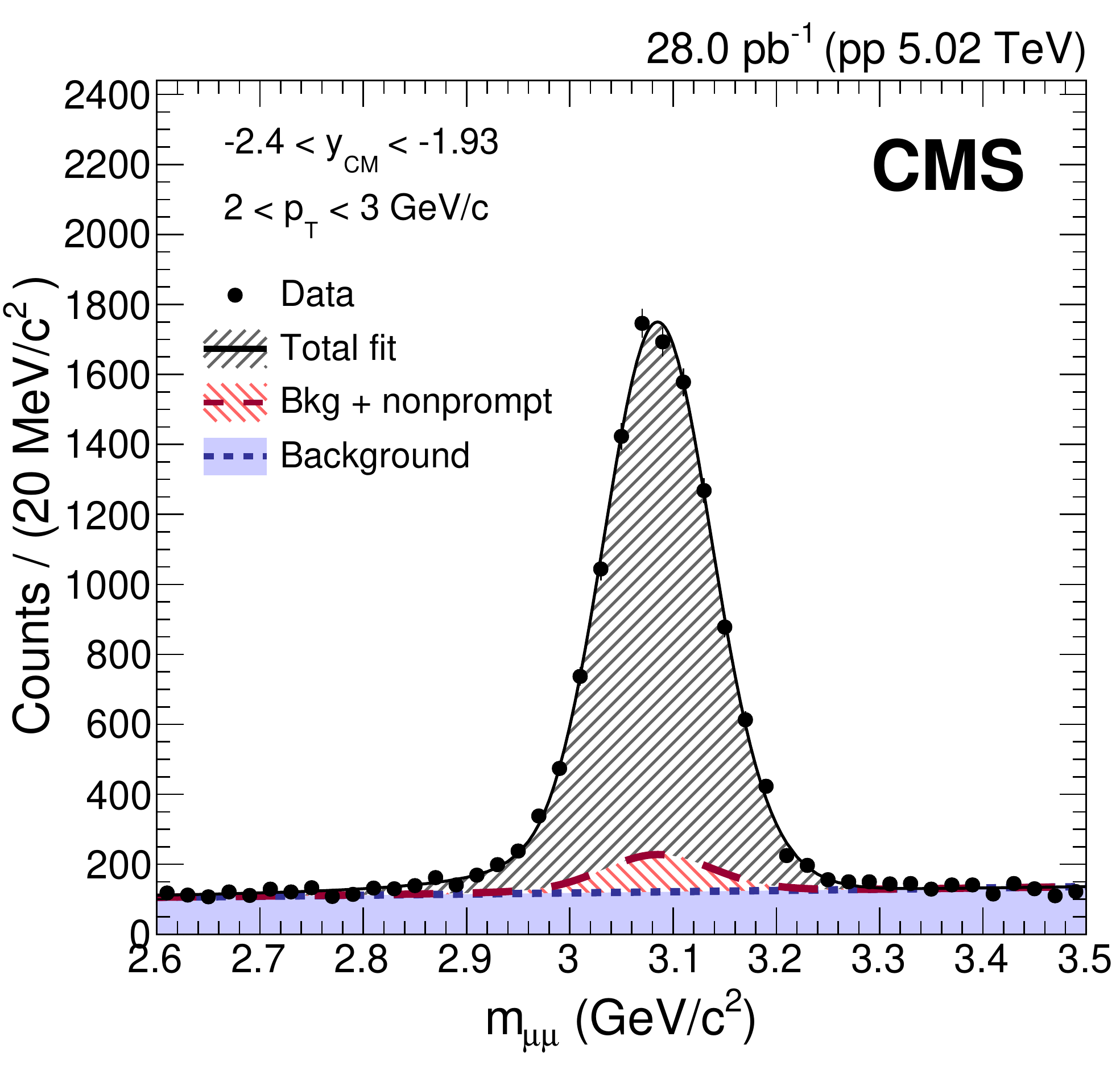}\hspace{1em}
\includegraphics[width=0.48\textwidth]{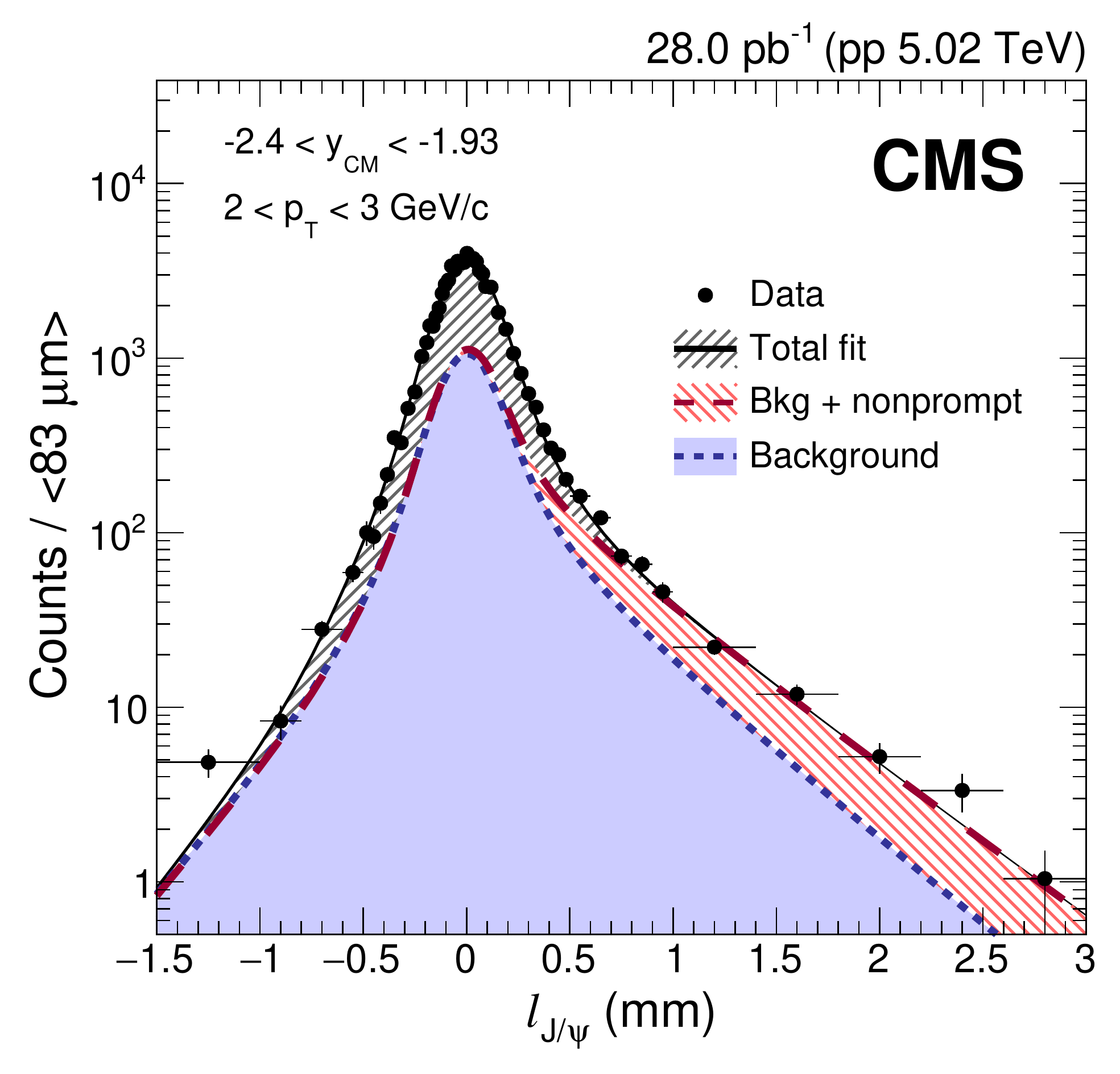}\\
\includegraphics[width=0.48\textwidth]{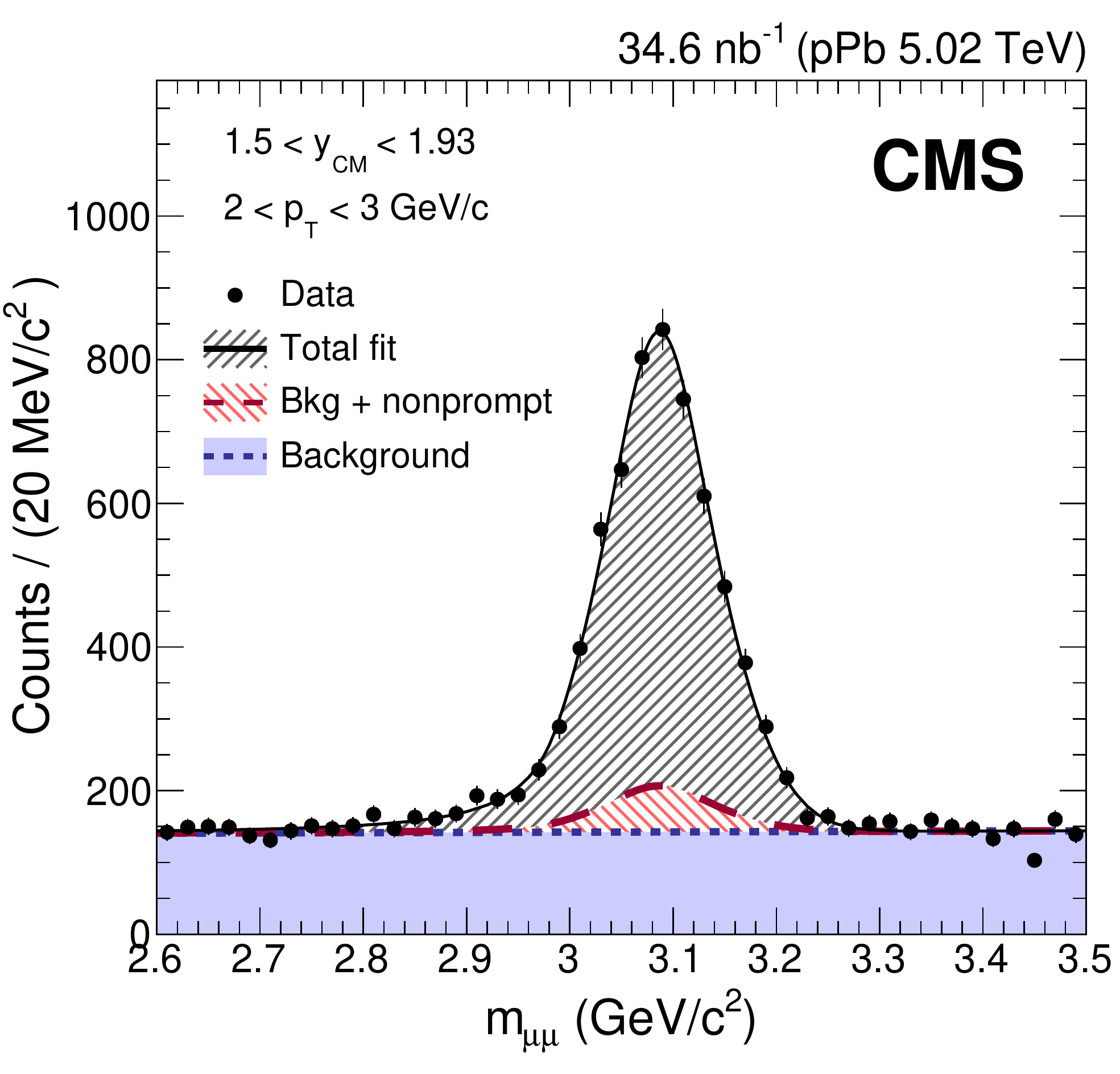}\hspace{1em}
\includegraphics[width=0.48\textwidth]{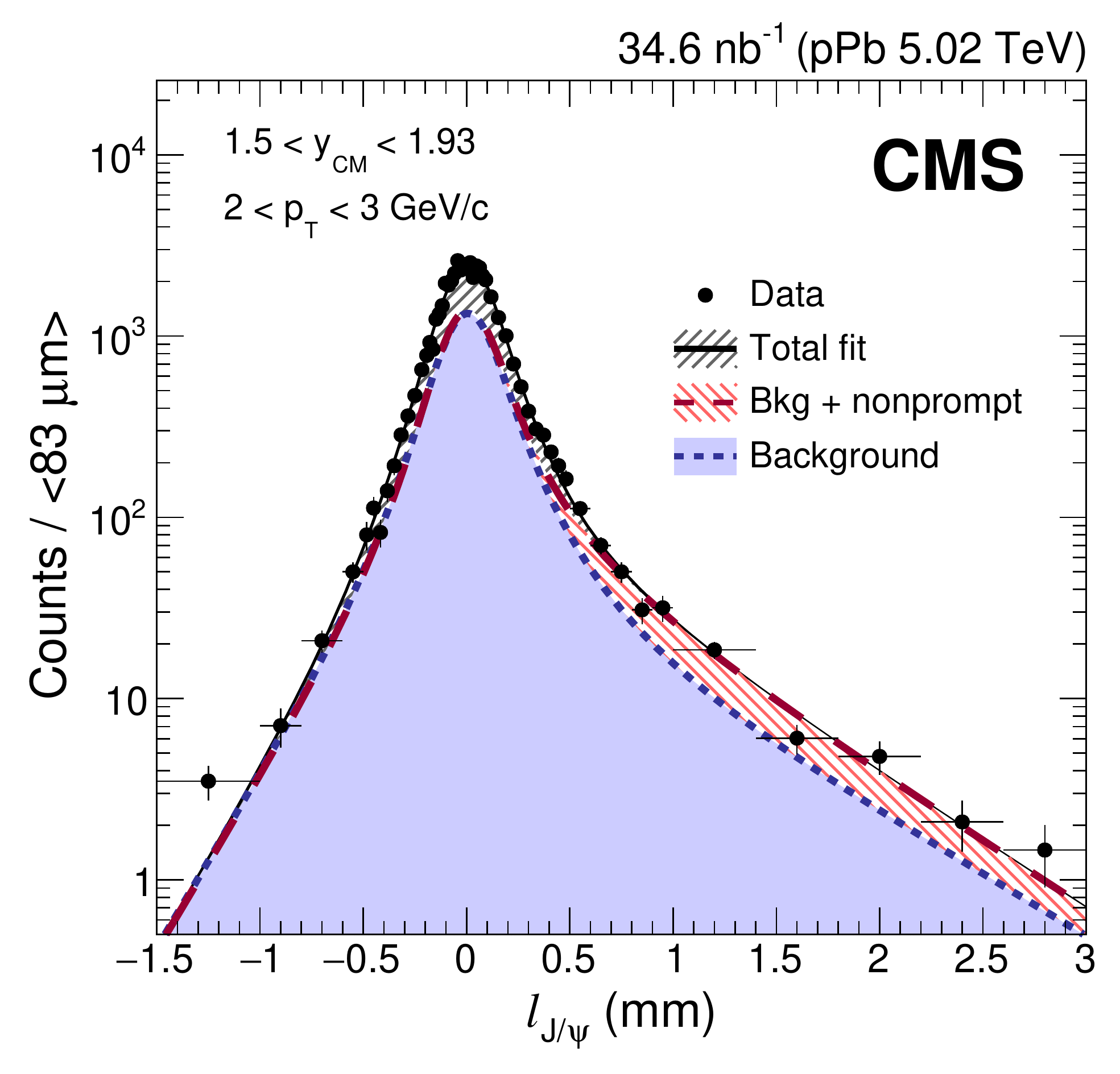}
\caption{Examples of the invariant mass (left) and pseudo-proper decay length (right) distributions of \mumu pairs for \pp (upper) and \pPb (lower) collisions. The bin widths of \ljpsi distributions vary from 15 to 500\unit{$\mu$m}, with the averaged value of 83\unit{$\mu$m}. The projections of the 2D fit function onto the respective axes are overlaid as solid lines. The long-dashed lines show the fitted contribution of nonprompt \JPsi mesons. The fitted background contributions are shown by short-dashed lines.}
\label{fig:jpsi_2dmassfits}
\end{center}
\end{figure*}

For the dimuon invariant mass distributions, the shape of the \JPsi signal is modeled by the sum of a Gaussian function and a Crystal Ball (CB) function~\cite{bib:CrystalBall}, with common mean values and independent widths, in order to accommodate the rapidity-dependent mass resolution. The CB function combines a Gaussian core with a power-law tail using two parameters \ncb and \acb, to describe final-state QED radiation of muons. Because the two parameters are strongly correlated, the value of \ncb is fixed at 2.1, while the \acb is a free parameter of the fit. This configuration gives the highest fit probability for data, in every $(\pt,\ycm)$ bin, when various settings of \acb and \ncb are tested. The invariant mass distribution of the underlying continuum background is represented by an exponential function.

For the \ljpsi distributions, the prompt signal component is represented by a resolution function, which depends on the per-event uncertainty in the \ljpsi provided by the reconstruction algorithm of primary and secondary vertices. The resolution function is composed of the sum of two Gaussian functions. A Gaussian with a narrower width (\sigN) describes the core of the signal component, while another with a greater width (\sigW) accounts for the effect of uncertainties in the primary vertex determination and has a fixed value based on MC simulations. The \ljpsi distribution of the nonprompt component is modeled by an exponential decay function convolved with a resolution function. The continuum background component is modeled by the sum of three exponential decay functions, a normal one on one side $\ljpsi>0$, a flipped one on the other side $\ljpsi<0$, and a double-sided one, which are also convolved with a resolution function. The parameters describing the \ljpsi distributions of the background are determined from sidebands in the invariant mass distribution $2.6<m_{\mu\mu}<2.9\GeVcc$ and $3.3<m_{\mu\mu}<3.5\GeVcc$. The results are insensitive to the selection of sideband ranges.

For \pPb analysis, two data sets corresponding to each beam direction are merged and fitted together, after it is determined that the results are compatible with those from a separate analysis, performed over each data set. Figure~\ref{fig:jpsi_2dmassfits} shows examples of fit projections onto the mass (left) and \ljpsi (right) axes for muon pairs with $2<\pt<3\GeVc$ in $-2.4<\ycm<-1.93$ from \pp (upper), and in $1.5<\ycm<1.93$ from \pPb (lower) collisions.

\subsection{Corrections}
\label{sec:corrections}

The acceptance and reconstruction, identification, and trigger efficiency corrections are evaluated from the MC simulation described in Section~\ref{sec:eventselection}. The acceptance is estimated by the fraction of generated \JPsi mesons in each $(\pt,\ycm)$ bin, decaying into two muons, each within the fiducial phase space defined in Eq.~(\ref{eqn:acccut}).

{\tolerance=1800
In order to compensate for imperfections in the simulation-based efficiencies, an additional scaling factor is applied, calculated with a \textit{tag-and-probe} (T\&P) method~\cite{Khachatryan:2010xn}. The tag muons require tight identification, and the probe muons are selected with and without satisfying the selection criteria relevant to the efficiency being measured. Then, invariant mass distributions of tag and probe pairs in the \JPsi mass range are fitted to count the number of signals in the two groups. The single-muon efficiencies are deduced from the ratio of \JPsi mesons in the passing-probe over all-probe group. The data-to-simulation ratios of single-muon efficiencies are used to correct the dimuon efficiencies, taking the kinematic distributions of decayed muons into account. The dimuon efficiency weights evaluated by the T\&P method are similar for \pp and \pPb events and range from 0.98 to 1.90, with the largest one coming from the lowest \pt bin. The efficiencies are independent of the event activity, as verified by \pPb data and in a \PYTHIA sample embedded in simulated \pPb events generated by \HIJING 1.383~\cite{Wang:1991hta}.
\par}

In addition, the shape of the uncorrected distributions of \JPsi yield versus \pt in data and MC samples are observed to be different. To resolve the possible bias in acceptance and efficiency corrections, the data-to-simulation ratios are fitted by empirical functions and used to reweight the \pt spectra in MC samples for each \ycm bin. The effect of reweighting on the acceptance and efficiency is detailed in the next Section.

\subsection{Systematic uncertainties}
\label{sec:systematicuncertainties}

The following sources of systematic uncertainties are considered: fitting procedure, acceptance and efficiency corrections, and integrated luminosities.

To estimate the systematic uncertainty due to the fitting procedure, variations of the parameters or alternative fit functions have been considered for the invariant mass and \ljpsi distributions. For the signal shape in the invariant mass distributions, three alternative parameter settings are tested: (1) \acb is set to 1.7, averaged from the default fit, and \ncb free, (2) both \acb and \ncb are left free, and (3) both are obtained from a MC template and then fixed when fit to the data. The maximum deviation of yields among these three variations is quoted as the uncertainty. For the background fit of the invariant mass distributions, a first-order polynomial is used as an alternative. For the shape of \ljpsi distribution of prompt \JPsi mesons, two alternatives are studied: (1) both \sigW and \sigN are left free, and (2) both parameters are fixed to the MC templates. The maximum deviation of yields is taken as the uncertainty. Finally, for the \ljpsi distribution shape of nonprompt \JPsi mesons, the template shape is directly taken from reconstructed MC events. The uncertainties from the previously mentioned methods are 0.7--5.0\% for prompt and 1.1--36.3\% for nonprompt \JPsi mesons. They are larger for the shape variations in the \ljpsi than in the invariant mass distributions, especially for nonprompt \JPsi mesons.

For the uncertainties from acceptance and efficiency correction factors, the effect of reweighting the \pt spectrum of events generated by \PYTHIA generator as described in Section~\ref{sec:corrections} is considered. The deviation of the correction factors obtained from the default \PYTHIA spectra and those from data-based weighted spectra is less than 2.9\% across all kinematic ranges. The full deviation values are quoted as the systematic uncertainties. The determination of uncertainties for T\&P corrections is performed by propagating the uncertainties in single-muon efficiencies to the dimuon efficiency values. The systematic uncertainties are evaluated by varying the fit conditions in the T\&P procedure, and the statistical uncertainties are estimated using a fast parametric simulation. The total uncertainty from T\&P corrections is obtained by the quadratic sum of two sources. Uncertainties from the efficiency correction, including the T\&P uncertainties, range from 2.4 to 6.1\%, and tend to be larger for lower \pt. The uncertainty in the integrated luminosities (2.3\% for \pp~\cite{CMS-PAS-LUM-16-001} and 3.5\% for \pPb~\cite{CMS-PAS-LUM-13-002}) is correlated across all data points and affects only the production cross sections and \rppb, while it cancels out in the \rfb measurements.

Table~\ref{tab:syst_tot} summarizes systematic uncertainties considered in this analysis. The range refers to different $(\pt,\ycm)$ bins; the uncertainties tend to be lower at high \pt and midrapidity, and higher at low \pt and forward or backward \ycm. The larger uncertainties of the nonprompt \JPsi yields come from the signal extraction in their lowest \pt bin, 2--3\GeVc. In the case of the \rppb measurements with a \pt limit of 4\GeVc, maximum uncertainties for nonprompt \JPsi mesons are 12.7\% for \pp and 12.8\% for \pPb collisions. The total systematic uncertainty is evaluated as the quadratic sum of the uncertainties from all sources in each kinematic bin, except for those from the integrated luminosity determination.

\begin{table*}[htbp]
\topcaption{Summary of the relative systematic uncertainties for the cross section measurements, given in percentages, for prompt and nonprompt \JPsi mesons in \pp and \pPb collisions.}
\label{tab:syst_tot}
\centering
\begin{tabular}{lcc{c}@{\hspace*{5pt}}cc}
\hline
\multirow{2}{*}{ } & \multicolumn{2}{c}{Prompt \JPsi} && \multicolumn{2}{c}{Nonprompt \JPsi} \\
\cline{2-3}\cline{5-6}
 & \pp & \pPb && \pp & \pPb \\
\hline
 Signal extraction & 0.8--3.2 & 0.7--5.0 && 2.0--36.3 & 1.1--29.5 \\
 Efficiency & 2.4--4.4 & 2.4--6.1 && 2.4--4.3 & 2.4--6.1 \\
 Acceptance & 0.0--2.3 & 0.0--1.2 && 0.0--1.3 & 0.0--1.3 \\
 Integrated luminosity & 2.3 & 3.5 && 2.3 & 3.5 \\
\hline
 Total & 2.7--5.3 & 2.8--7.1 && 3.4--36.5 & 3.3--30.1 \\
\hline
\end{tabular}
\end{table*}

\section{Results}
\label{sec:results}

\subsection{Prompt \texorpdfstring{\JPsi}{J/psi} mesons}
\label{sec:promptjpsi}

\begin{figure*}[hbtp]
\centering
\includegraphics[width=0.48\textwidth]{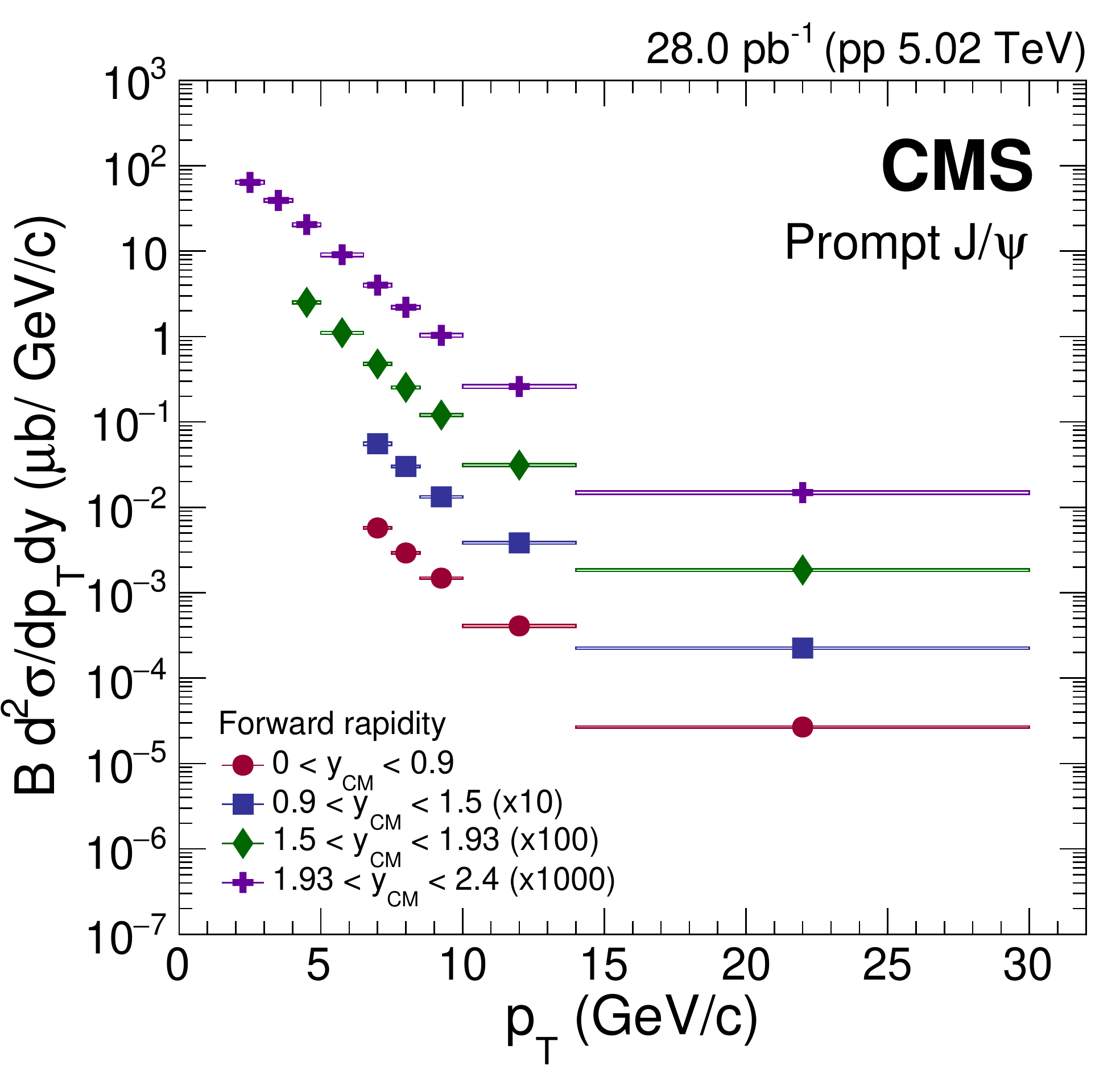}
\includegraphics[width=0.48\textwidth]{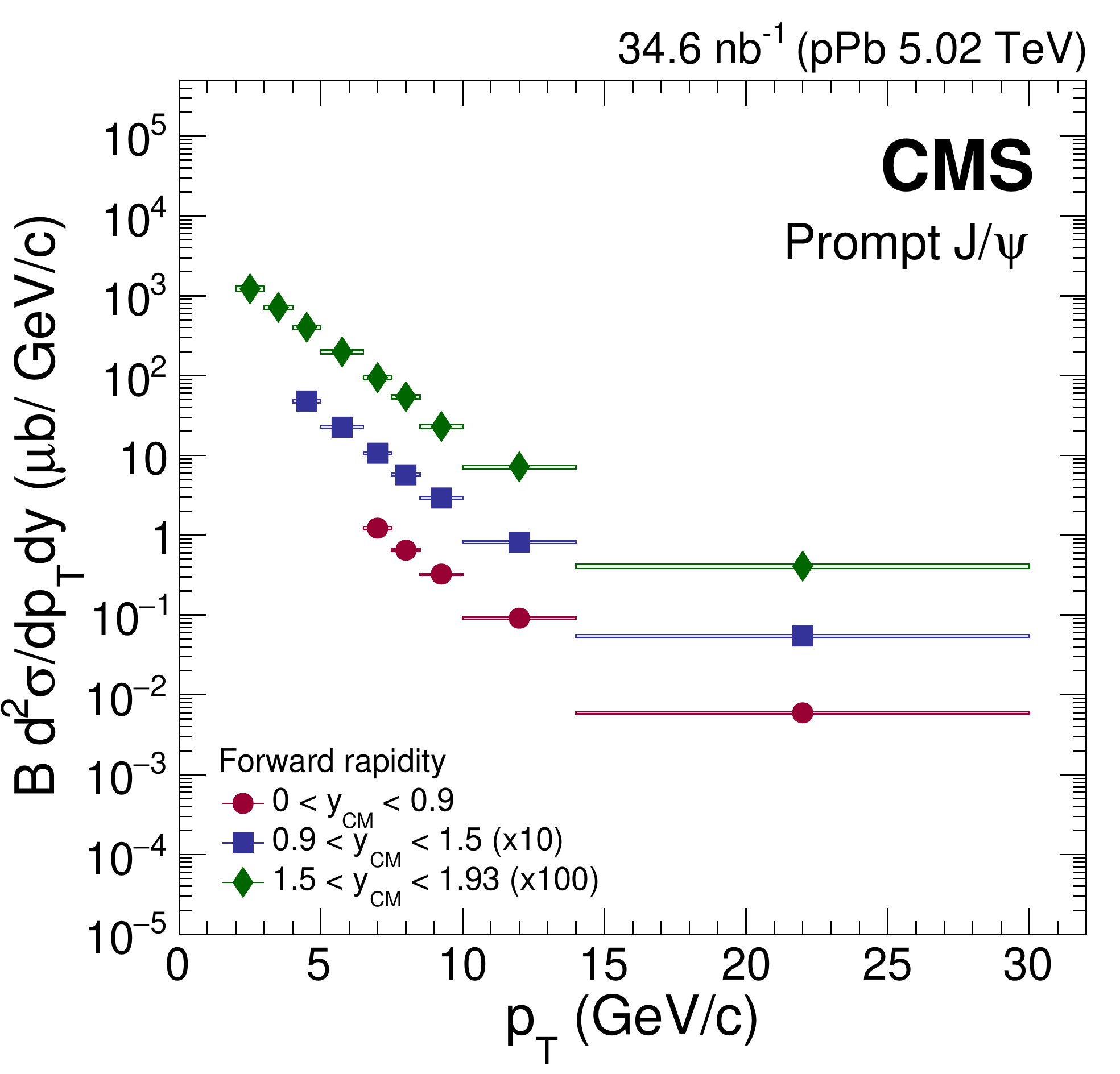}
\includegraphics[width=0.48\textwidth]{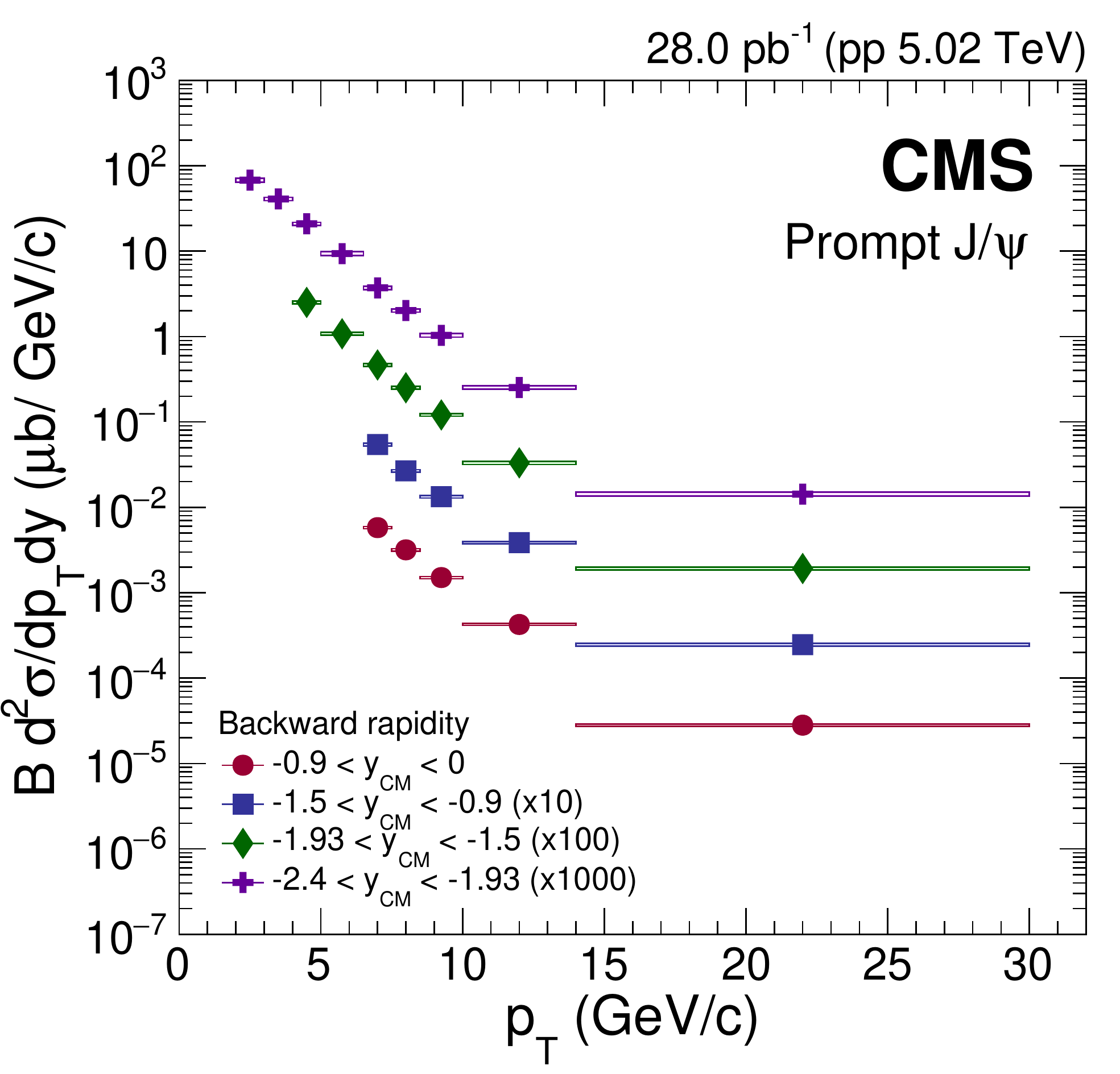}
\includegraphics[width=0.48\textwidth]{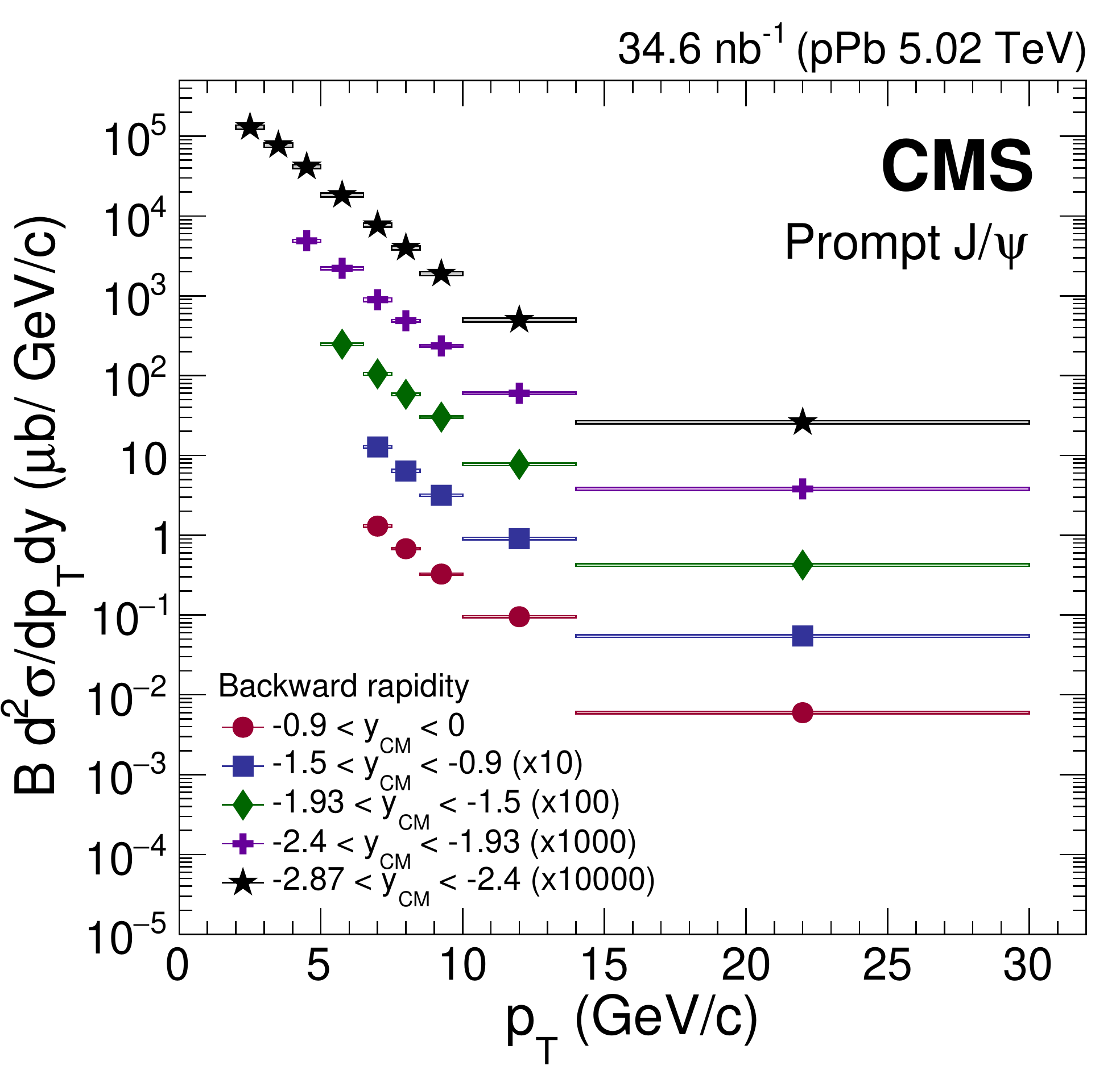}
\caption{Differential cross section (multiplied by the dimuon branching fraction) of prompt \JPsi mesons in \pp (left) and \pPb (right) collisions at forward (upper) and backward (lower) \ycm. The vertical bars (smaller than the symbols in most cases) represent the statistical uncertainties and the shaded boxes show the systematic uncertainties. The fully correlated global uncertainty from the integrated luminosity determination, 2.3\% for \pp and 3.5\% for \pPb collisions, is not included in the point-by-point uncertainties.}
\label{fig:cross_pt_pr}
\end{figure*}

\begin{figure}[hbtp]
\centering
\includegraphics[width=0.48\textwidth]{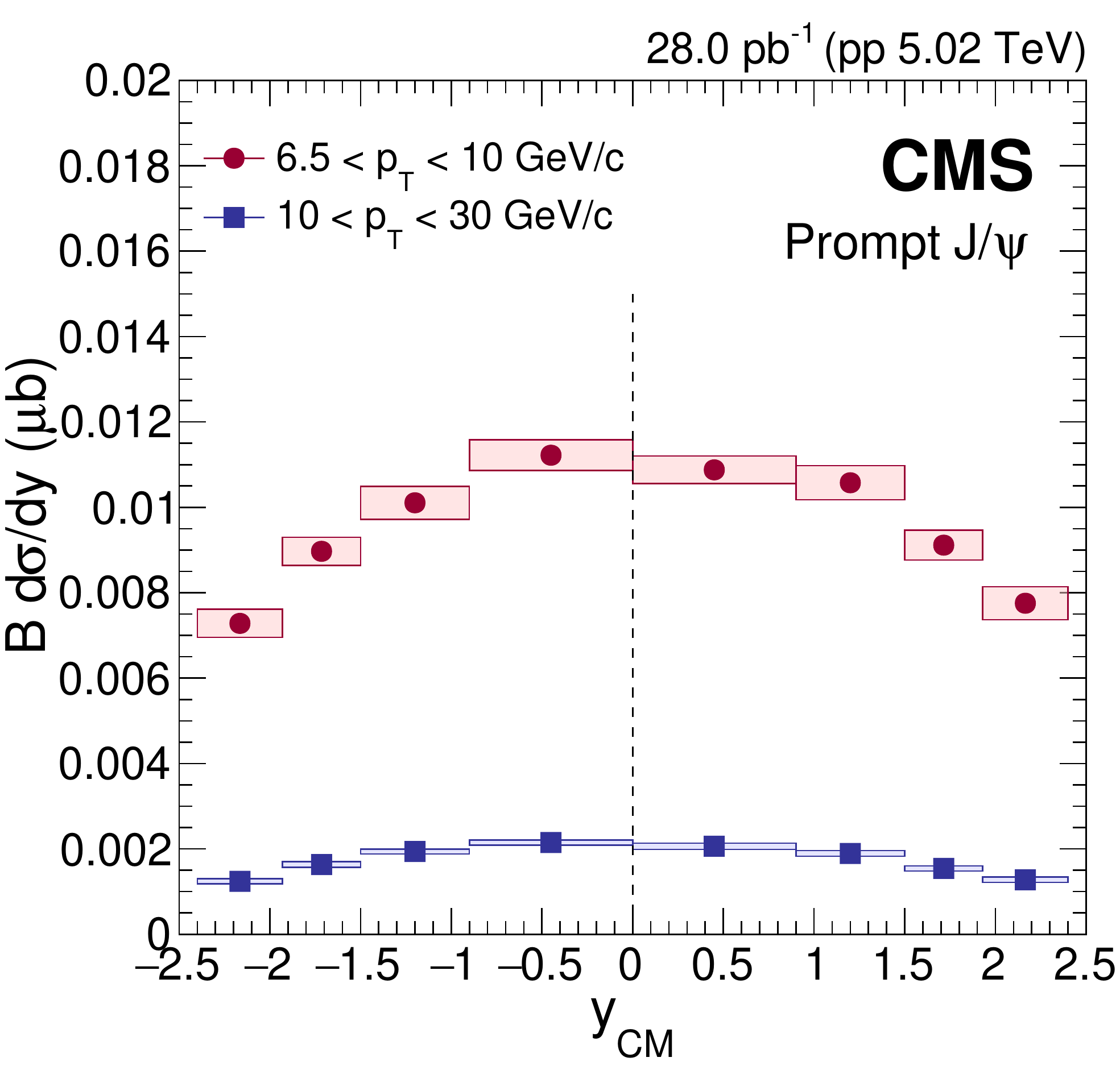}
\includegraphics[width=0.48\textwidth]{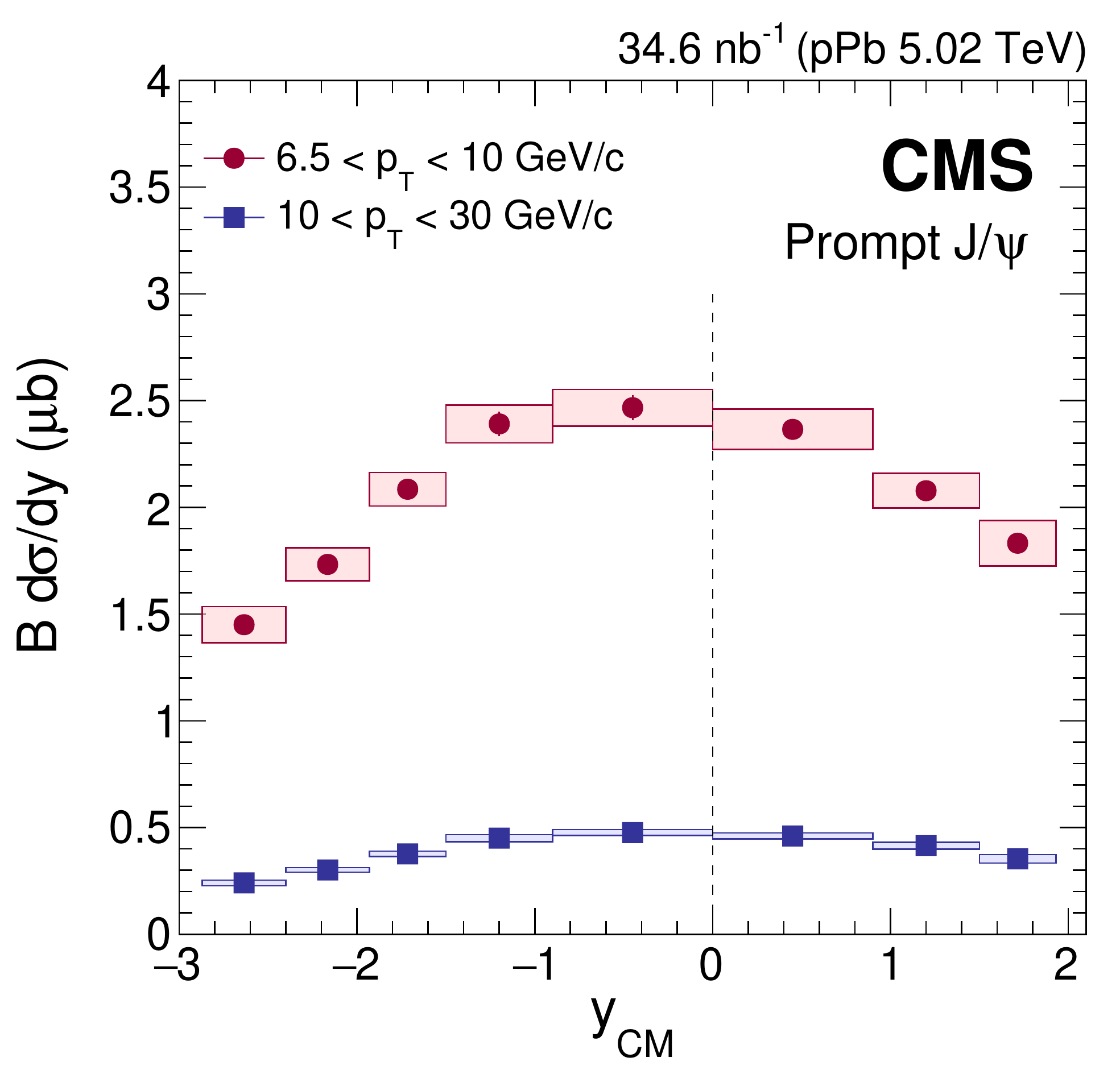}
\caption{Rapidity dependence of the cross section (multiplied by the dimuon branching fraction) for prompt \JPsi mesons in the \pt intervals of $6.5<\pt<10\GeVc$ (circles) and $10<\pt<30\GeVc$ (squares) in \pp (\cmsLeft) and \pPb (\cmsRight) collisions. The vertical dashed line indicates $\ycm=0$. The vertical bars (smaller than the symbols in most cases) represent the statistical uncertainties and the shaded boxes show the systematic uncertainties. The fully correlated global uncertainty from the integrated luminosity determination, 2.3\% for \pp and 3.5\% for \pPb collisions, is not included in the point-by-point uncertainties.}
\label{fig:cross_rap_pr}
\end{figure}

Figure~\ref{fig:cross_pt_pr} shows the double-differential prompt \JPsi production cross sections multiplied by the dimuon branching fraction in \pp (left) and \pPb (right) collisions, with data points plotted at the center of each bin. Statistical uncertainties are displayed as vertical bars, while boxes that span the \pt bin width represent systematic uncertainties. Not shown is a global normalization uncertainty of 2.3\% in \pp and 3.5\% in \pPb collisions arising from the integrated luminosity determination.

Prompt \JPsi \ycm distributions are shown in Fig.~\ref{fig:cross_rap_pr} in \pp (\cmsLeft) and \pPb (\cmsRight) collisions. The measurements are integrated over two \pt intervals, $6.5<\pt<10\GeVc$ (low \pt) and $10<\pt<30\GeVc$ (high \pt).

\begin{figure*}[hbtp]
\centering
\includegraphics[width=0.98\textwidth]{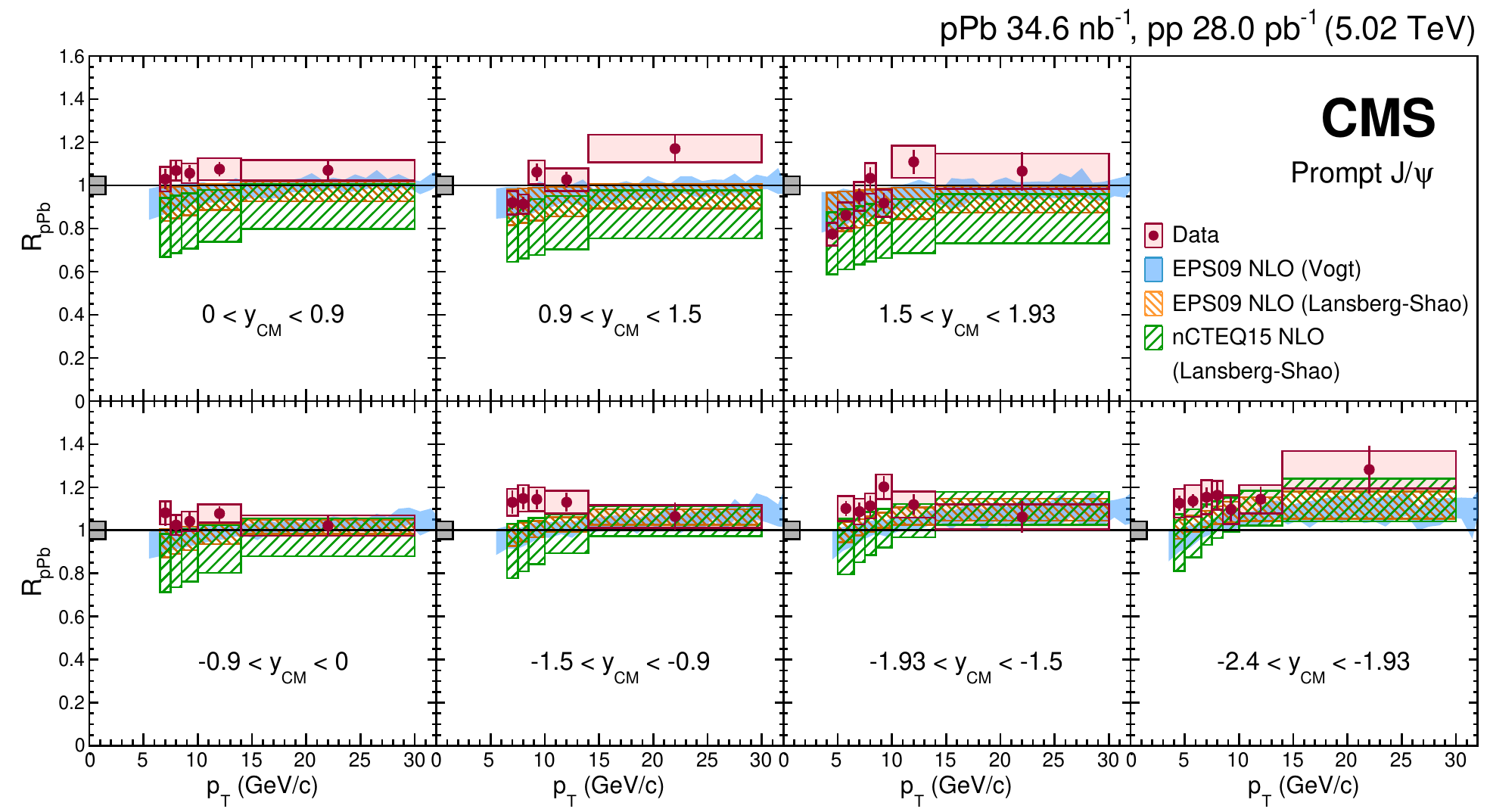}
\caption{Transverse momentum dependence of \rppb for prompt \JPsi mesons in seven \ycm ranges. The vertical bars represent the statistical uncertainties and the shaded boxes show the systematic uncertainties. The fully correlated global uncertainty of 4.2\% is displayed as a gray box at $\rppb=1$ next to the left axis. The predictions of shadowing models based on the parameterizations EPS09 and nCTEQ15~\cite{Vogt:2015uba,Eskola:2009uj,PhysRevD.93.085037,Lansberg:2016deg} are also shown.}
\label{fig:rppb_pt_pr}
\end{figure*}

The \pt dependence of prompt \JPsi \rppb is shown in Fig.~\ref{fig:rppb_pt_pr}, in seven \ycm ranges for which \pp and \pPb measurements overlap. Around midrapidity ($\abs{\ycm}<0.9$) and in the three backward \ycm bins (lower panels), \rppb is slightly above unity without a clear dependence on \pt. In the most forward bin ($1.5<\ycm<1.93$), suppression at low \pt ($\lesssim7.5\GeVc$) is observed, followed by a weak increase of \rppb at higher \pt. The results are compared to three model calculations. One is based on the next-to-leading order (NLO) Color Evaporation Model~\cite{Vogt:2015uba} using the EPS09~\cite{Eskola:2009uj} nPDF set. The other two are calculated from the nPDF sets of EPS09 and nCTEQ15~\cite{PhysRevD.93.085037}, respectively, with the parameterization of $2\to2$ partonic scattering process based on data, as described in Ref.~\cite{Lansberg:2016deg}. All three \rppb calculations are marginally lower than the measured values across all \ycm bins. The calculations based on coherent energy loss are not yet available to describe quarkonium production at large \pt ($\gtrsim m_{\JPsi}$); therefore, no comparison of the present data with the model~\cite{Arleo:2012hn} is performed.

It is worth noting that the \rppb values measured in the most forward ($1.5<\ycm<1.93$) and backward ($-2.4<\ycm<-1.93$) regions are consistent, in the overlapping \pt intervals ($4<\pt<8\GeVc$), with the inclusive \JPsi results of the ALICE collaboration~\cite{Abelev:2013yxa,Adam:2015iga} over $2.03<\ycm<3.53$ and $-4.46<\ycm<-2.96$, obtained using an interpolated \pp cross section reference. Although the ALICE results are for inclusive \JPsi mesons, the nonprompt contribution is expected to be relatively small ($<20\%$) in the domain $\pt<8\GeVc$.

\begin{figure}[hbtp]
\centering
\includegraphics[width=0.48\textwidth]{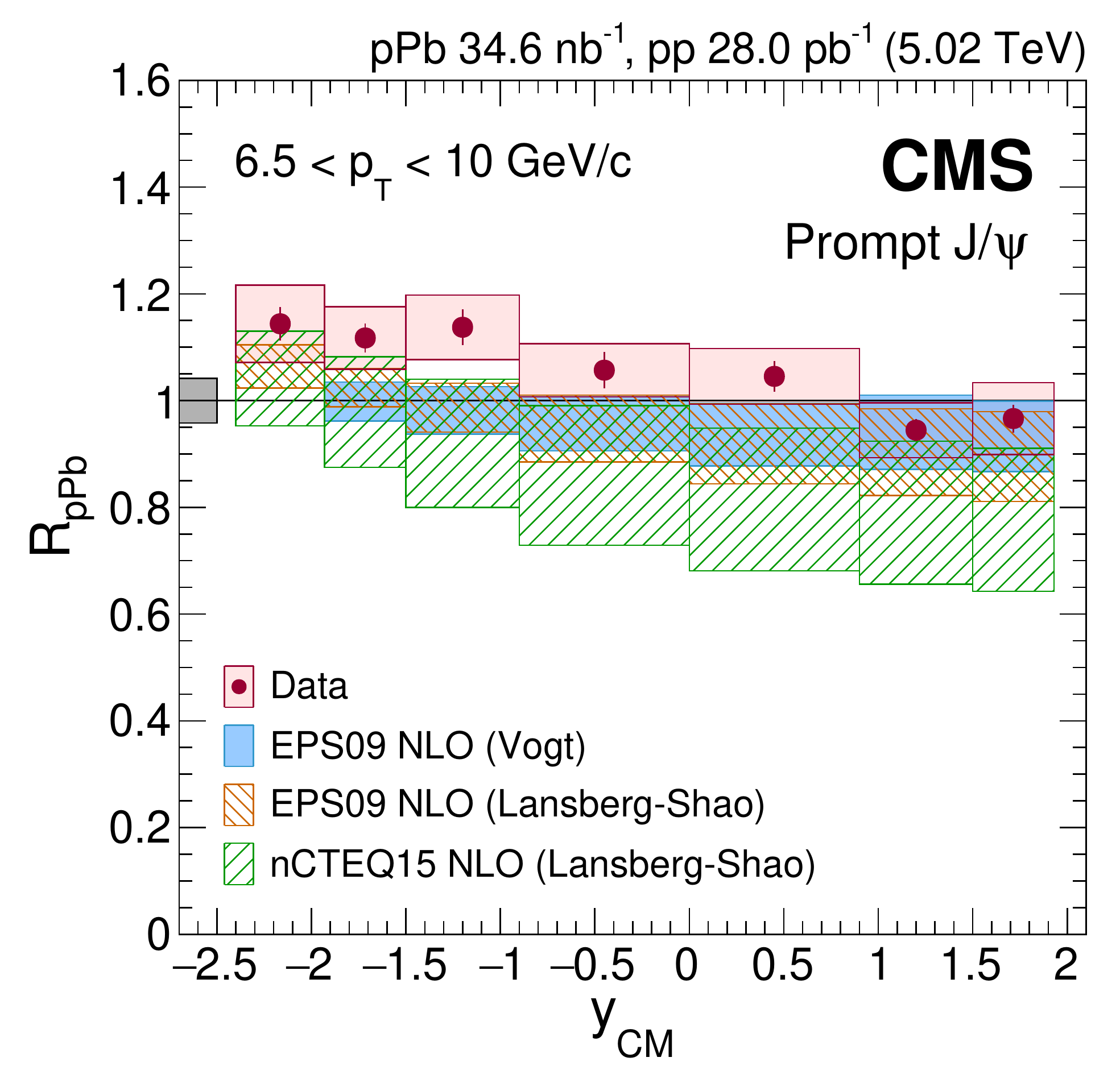}
\includegraphics[width=0.48\textwidth]{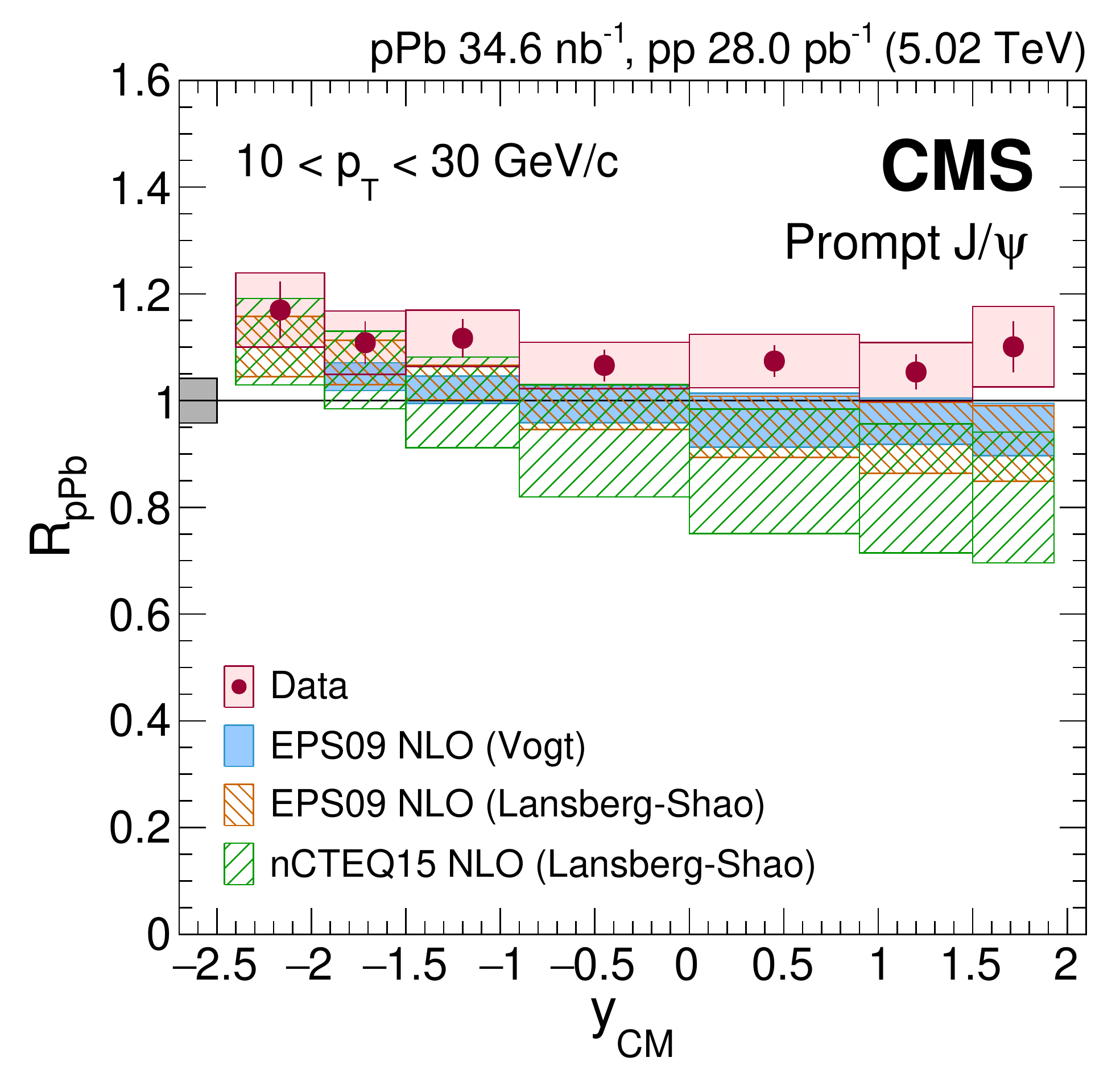}
\caption{Rapidity dependence of \rppb for prompt \JPsi mesons in two \pt ranges: $6.5<\pt<10\GeVc$ (\cmsLeft) and $10<\pt<30\GeVc$ (\cmsRight). The vertical bars represent the statistical uncertainties and the shaded boxes show the systematic uncertainties. The fully correlated global uncertainty of 4.2\% is displayed as a gray box at $\rppb=1$ next to the left axis. The predictions of shadowing models based on the parameterizations EPS09 and nCTEQ15~\cite{Vogt:2015uba,Eskola:2009uj,PhysRevD.93.085037,Lansberg:2016deg} are also shown.}
\label{fig:rppb_rap_pr}
\end{figure}

Figure~\ref{fig:rppb_rap_pr} displays the \ycm dependence of prompt \JPsi \rppb in the low-\pt (\cmsLeft) and the high-\pt (\cmsRight) regions corresponding to the same \pt bins used in Fig.~\ref{fig:cross_rap_pr}. In the high-\pt region, \rppb is above unity over the whole \ycm range. In the lower-\pt region, a decrease of \rppb for increasing $\ycm$ is suggested. The same theoretical predictions shown in Fig.~\ref{fig:rppb_pt_pr} are overlaid. In contrast to the measurement of \JPsi mesons in \PbPb collisions~\cite{Khachatryan:2016ypw}, no significant deviation from unity is observed in the \pt and \ycm ranges studied here. This suggests that the strong suppression of \JPsi production in \PbPb collisions is an effect of QGP formation.

\begin{figure}[hbtp]
\centering
\includegraphics[width=0.32\textwidth]{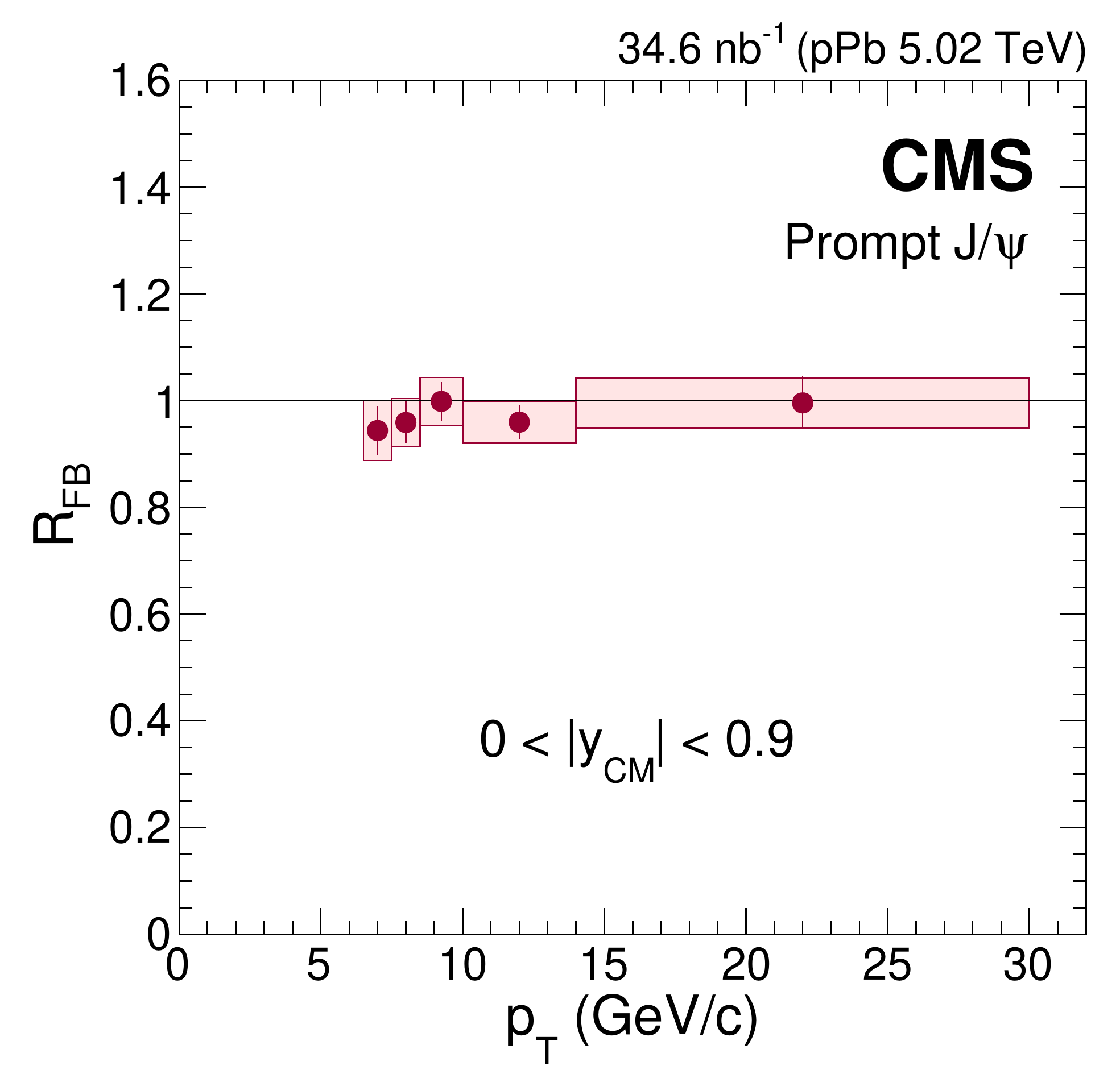}
\includegraphics[width=0.32\textwidth]{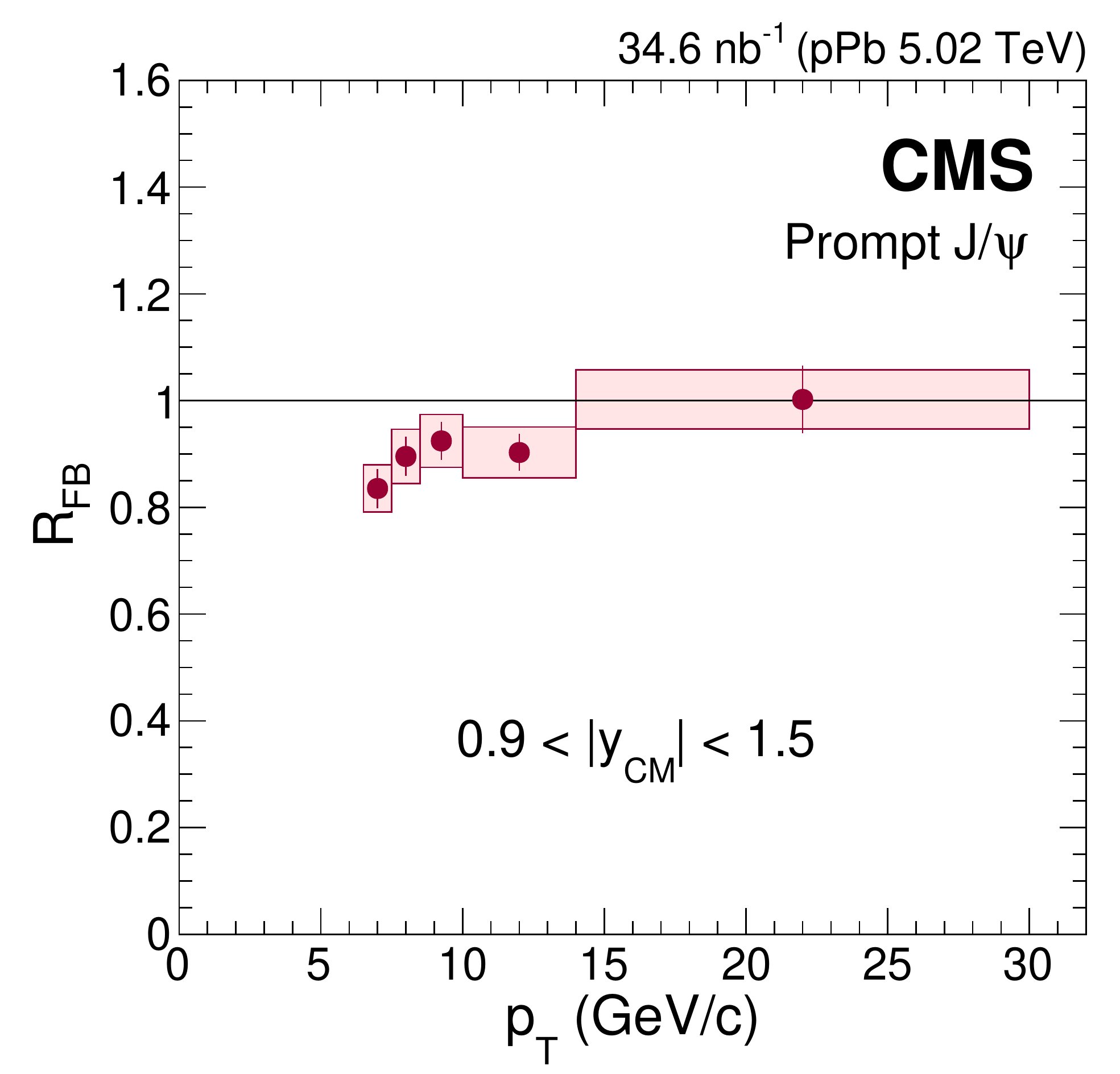}
\includegraphics[width=0.32\textwidth]{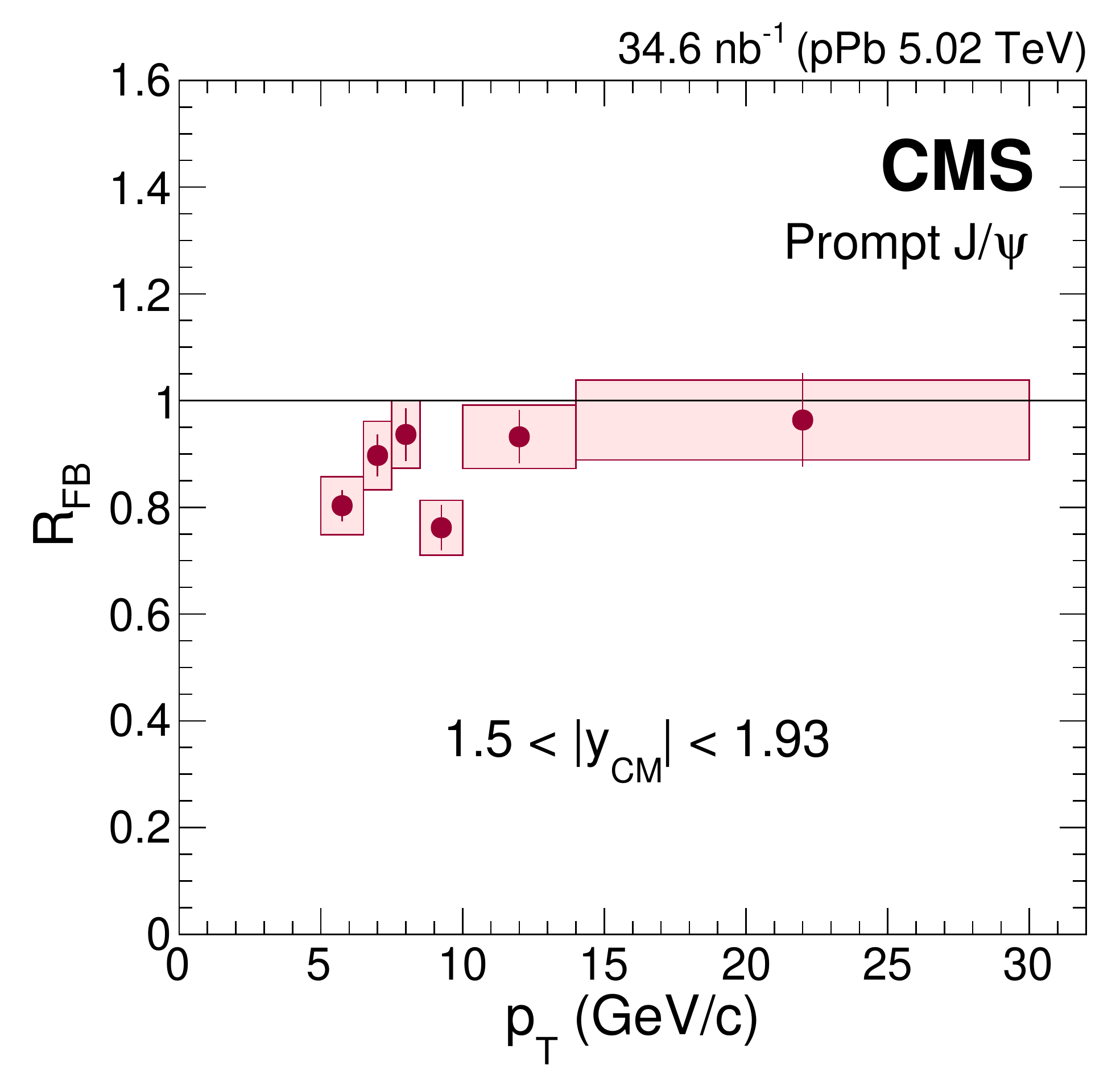}
\caption{Transverse momentum dependence of \rfb for prompt \JPsi mesons in three \ycm regions. The vertical bars represent the statistical uncertainties and the shaded boxes show the systematic uncertainties.}
\label{fig:rfb_pt_pr}
\end{figure}

The forward-to-backward ratio of \pPb cross sections, \rfb, in three \ycm ranges is displayed as a function of \pt for prompt \JPsi mesons in Fig.~\ref{fig:rfb_pt_pr}. The \rfb tends to be below unity at low $\pt\lesssim7.5\GeVc$ and forward $\abs{\ycm}>0.9$. In the $6.5<\pt<10\GeVc$ bin, an indication of decrease of \rfb with increasing \ycm is observed. The results are in agreement with the measurements from the ATLAS~\cite{Aad:2015ddl}, ALICE~\cite{Abelev:2013yxa,Adam:2015iga}, and LHCb~\cite{Aaij:2013zxa} collaborations.

\begin{figure}[hbtp]
\centering
\includegraphics[width=0.48\textwidth]{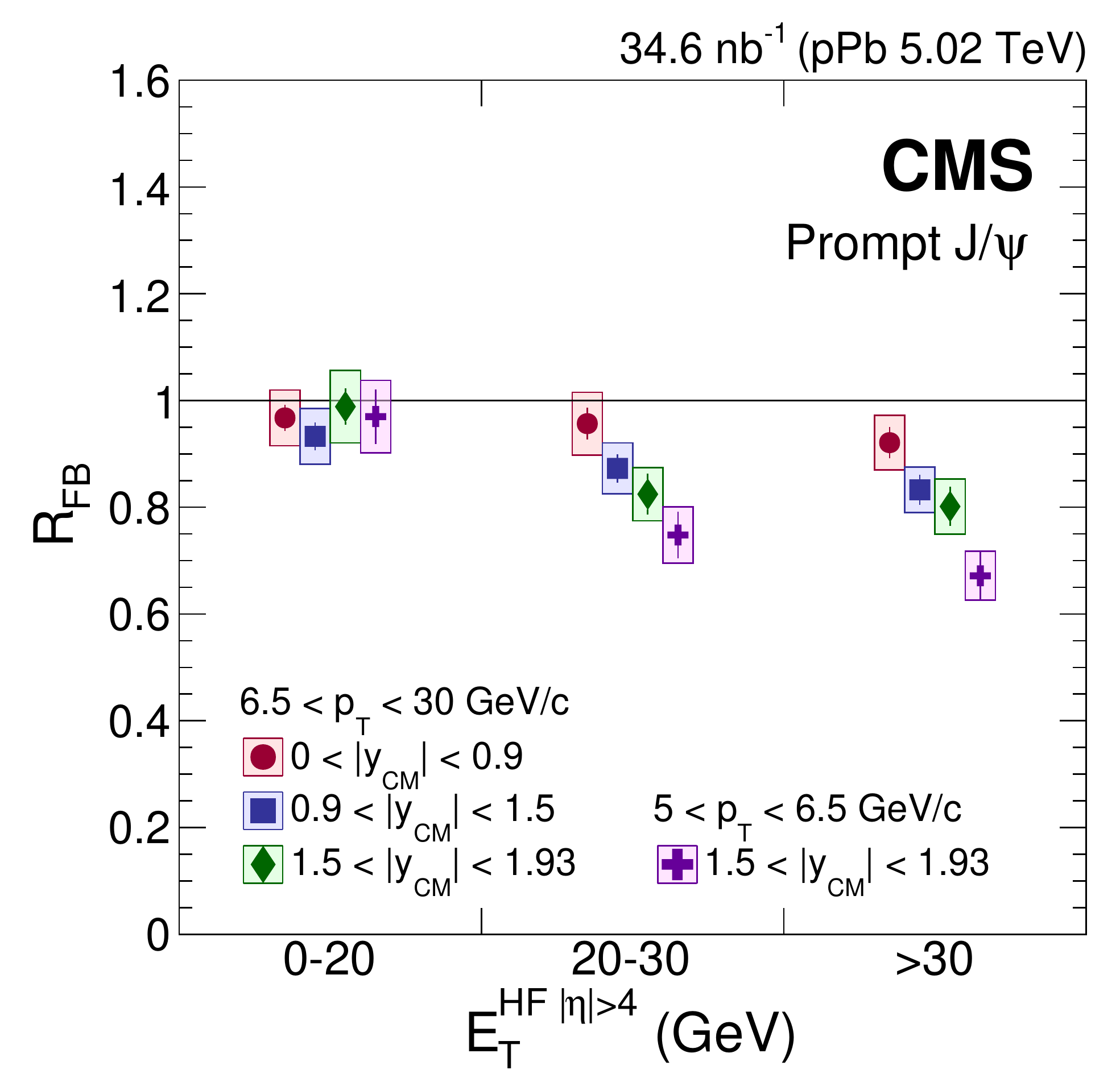}
\caption{Dependence of \rfb for prompt \JPsi mesons on the hadronic activity in the event, given by the transverse energy deposited in the CMS detector at large pseudorapidities \ethf. Data points are slightly shifted horizontally so that they do not overlap. The vertical bars represent the statistical uncertainties and the shaded boxes show the systematic uncertainties.}
\label{fig:rfb_ethf_pr}
\end{figure}

{\tolerance=1200
Figure~\ref{fig:rfb_ethf_pr} shows \rfb as a function of \ethf for prompt \JPsi mesons in three \ycm ranges. The data are integrated over $6.5<\pt<30\GeVc$; a lower-\pt bin, $5<\pt<6.5\GeVc$, is shown in addition for the most forward-backward interval, $1.5<\abs{\ycm}<1.93$. The value of \rfb decreases as a function of \ethf, suggesting that the effects that cause the asymmetry between the forward-to-backward production are larger in events with more hadronic activity.
\par}

\subsection{Nonprompt \texorpdfstring{\JPsi}{J/psi} mesons}
\label{sec:nonpromptjpsi}

\begin{figure*}[hbtp]
\centering
\includegraphics[width=0.48\textwidth]{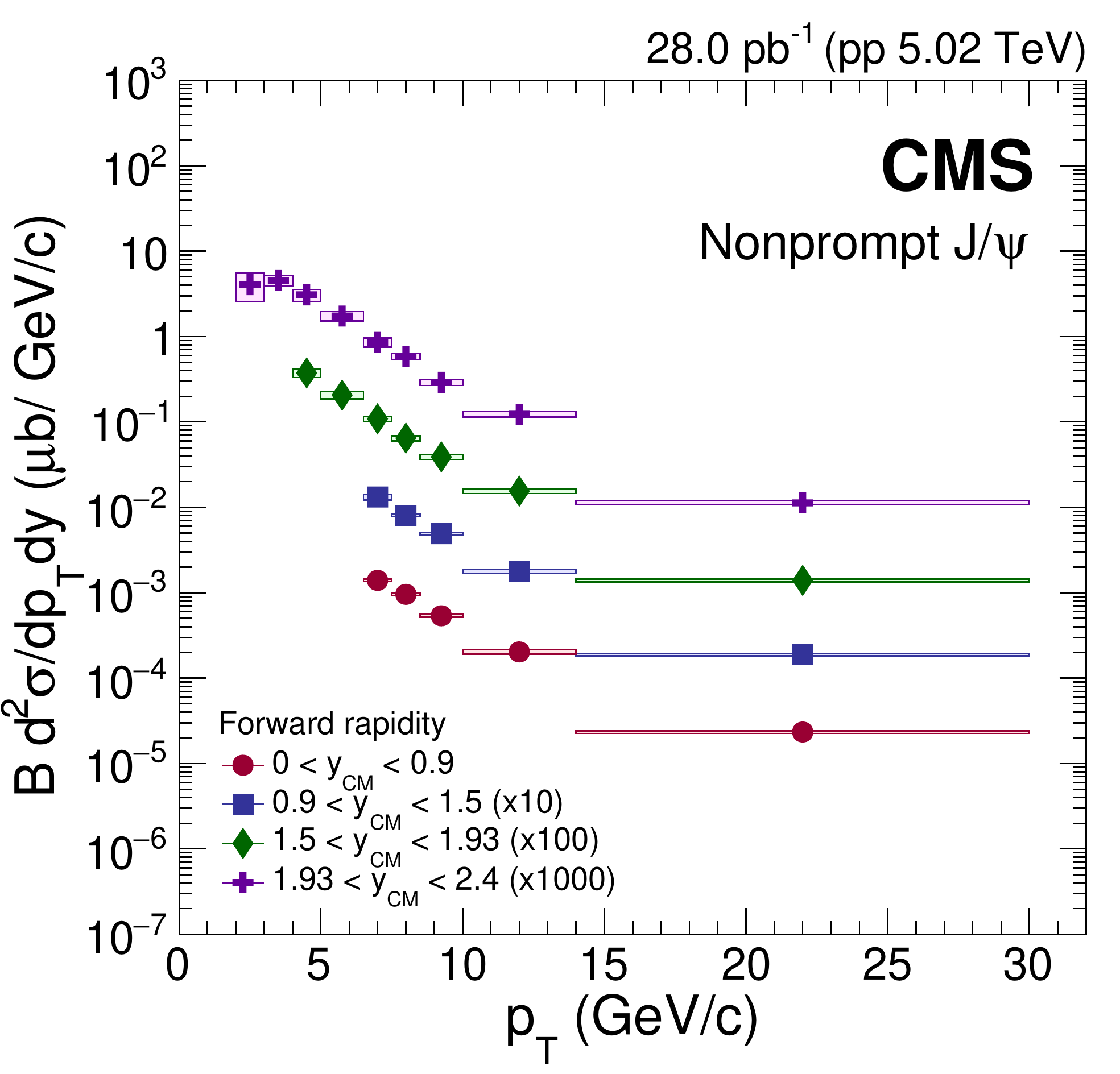}
\includegraphics[width=0.48\textwidth]{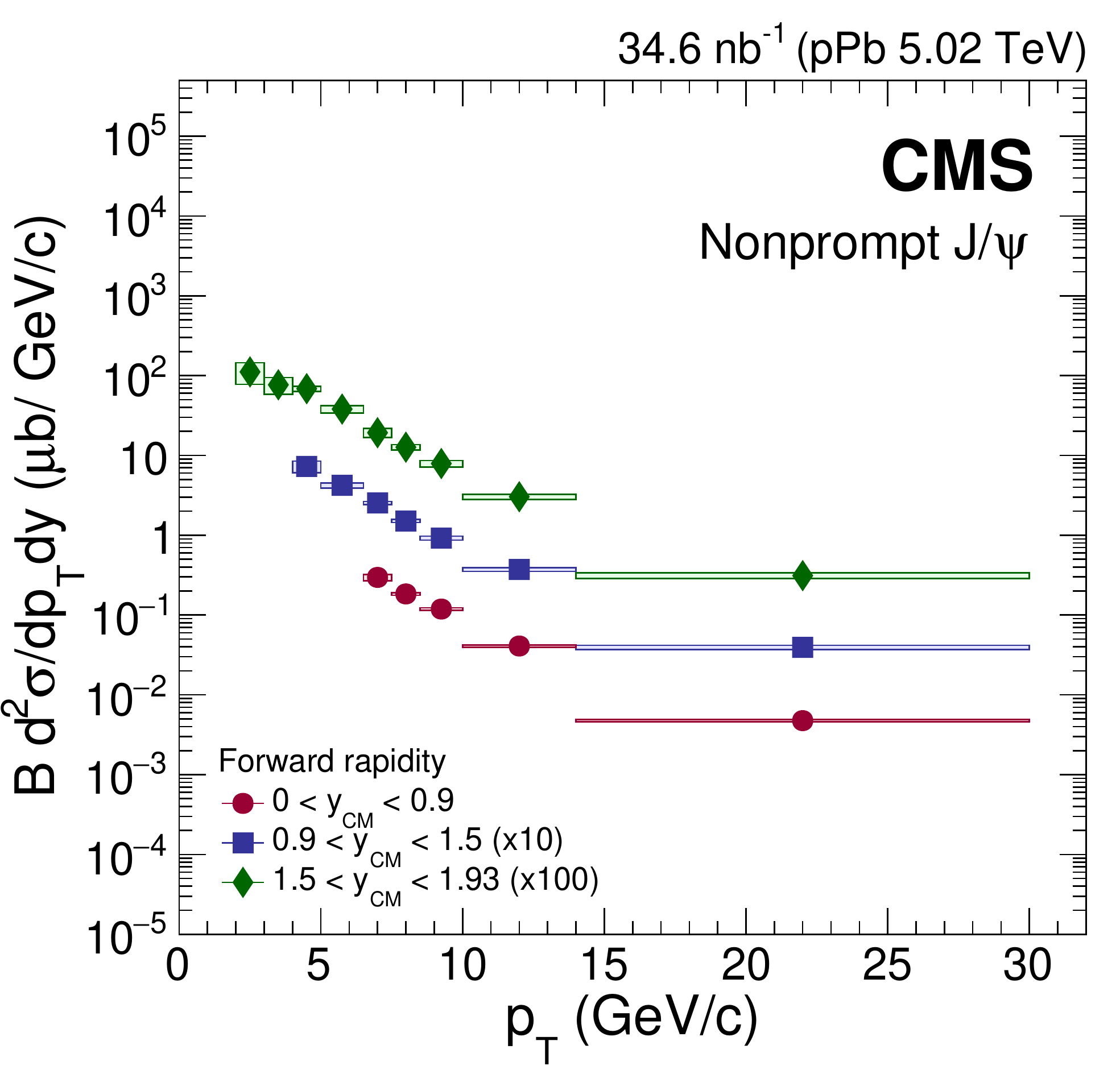}
\includegraphics[width=0.48\textwidth]{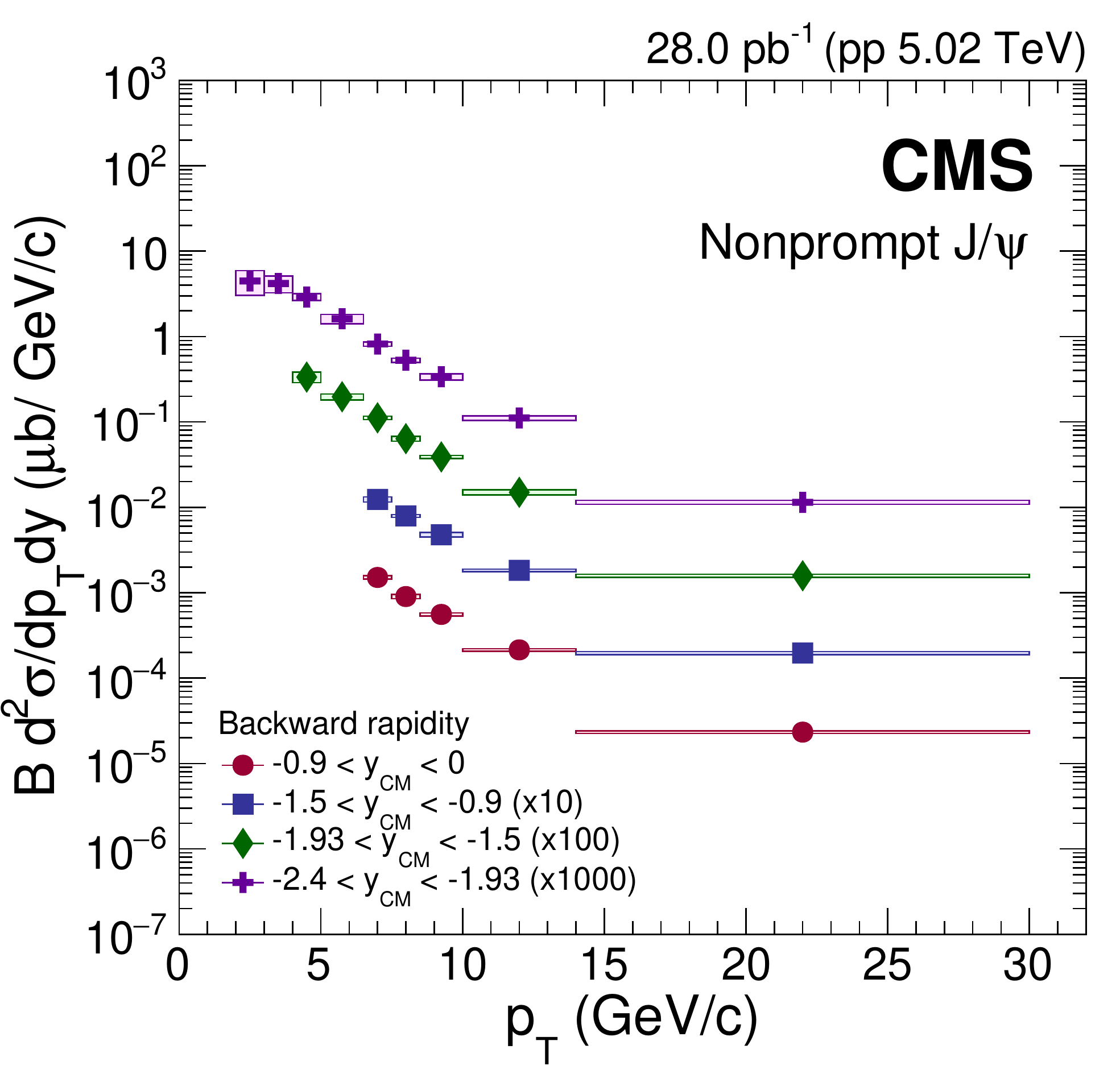}
\includegraphics[width=0.48\textwidth]{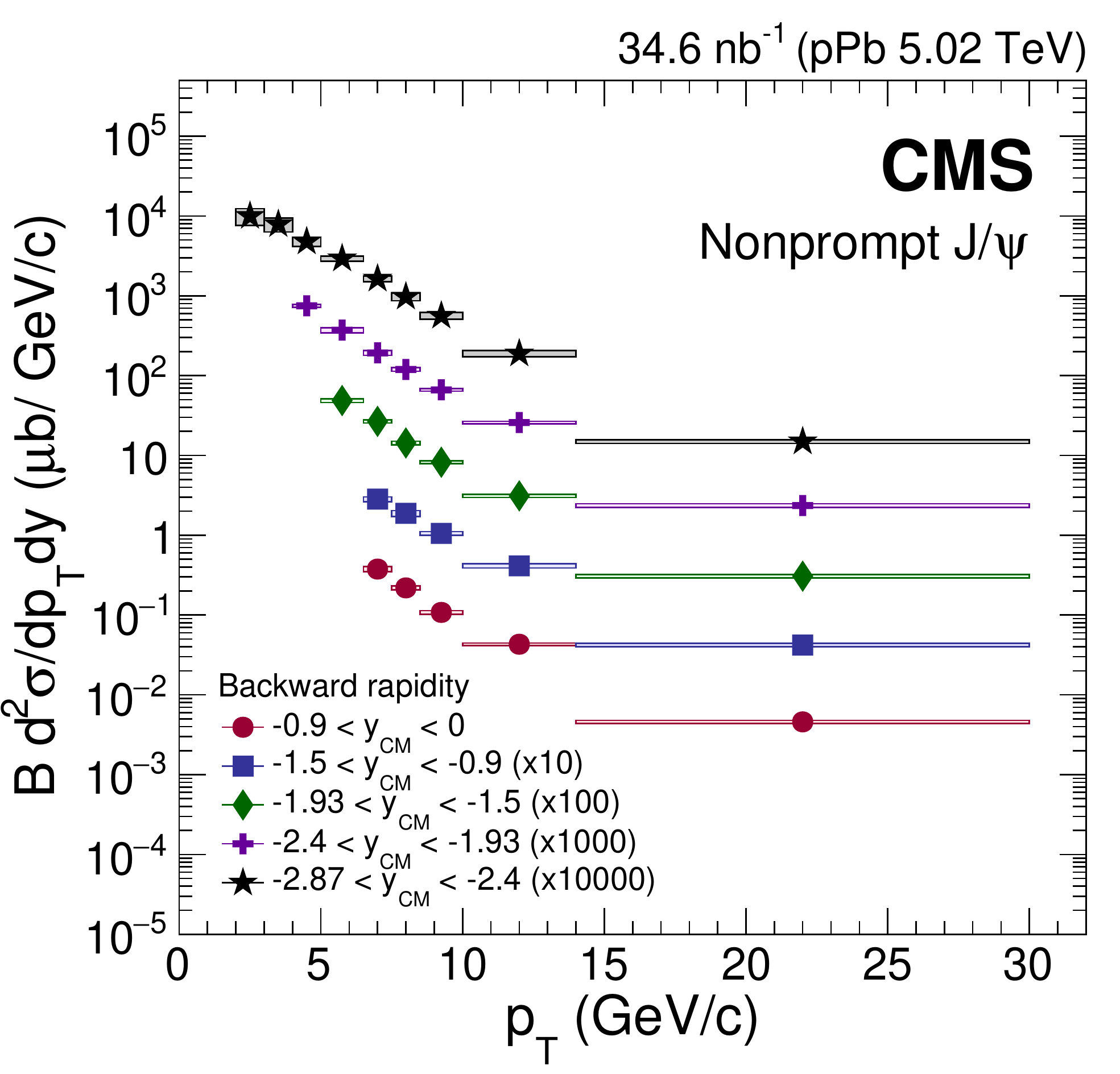}
\caption{Differential cross section (multiplied by the dimuon branching fraction) of nonprompt \JPsi mesons in \pp (left) and \pPb (right) collisions at forward (upper) and backward (lower) \ycm. The vertical bars (smaller than the symbols in most cases) represent the statistical uncertainties and the shaded boxes show the systematic uncertainties. The fully correlated global uncertainty from the integrated luminosity determination, 2.3\% for \pp and 3.5\% for \pPb collisions, is not included in the point-by-point uncertainties.}
\label{fig:cross_pt_np}
\end{figure*}

\begin{figure}[hbtp]
\centering
\includegraphics[width=0.48\textwidth]{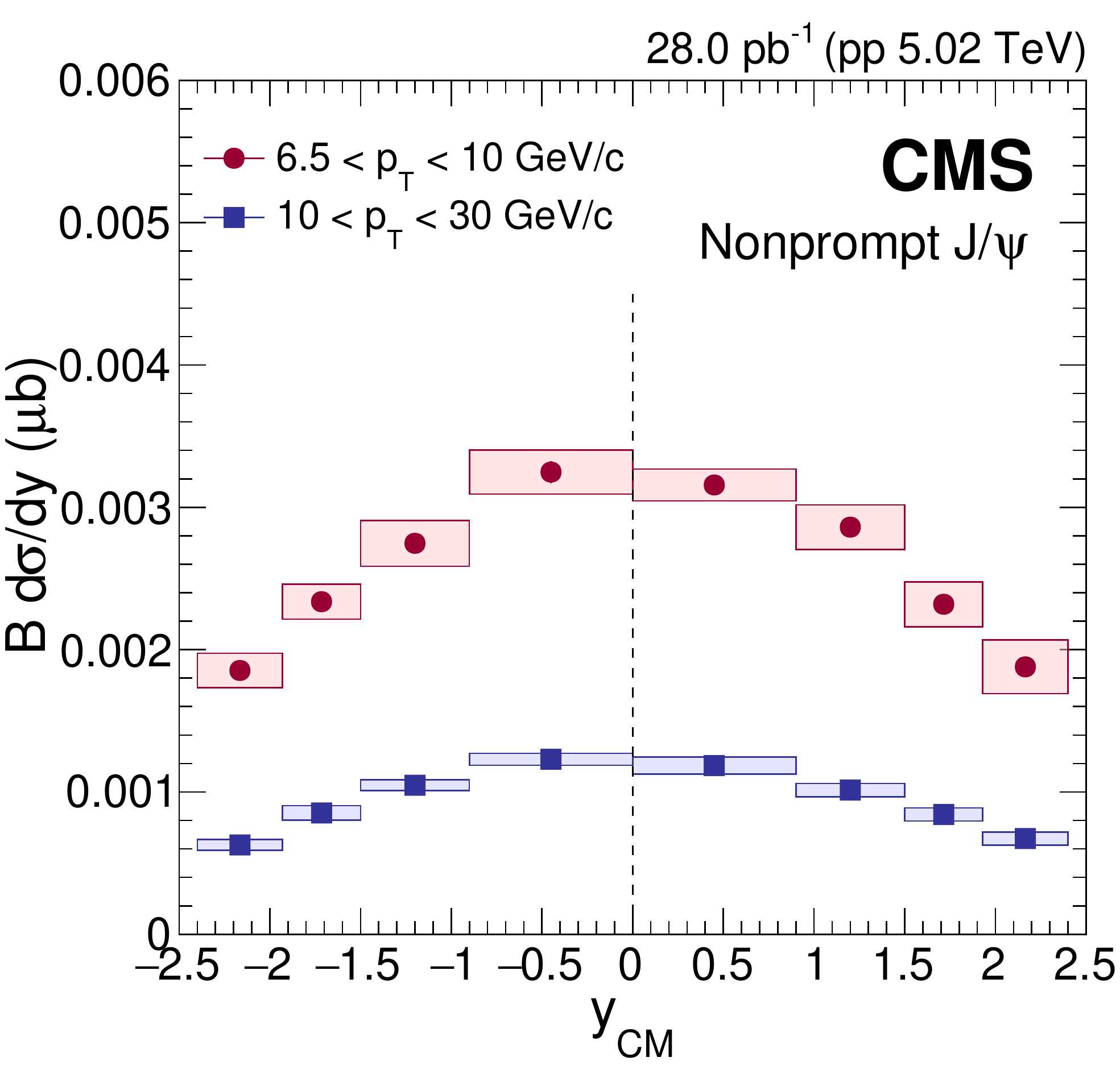}
\includegraphics[width=0.48\textwidth]{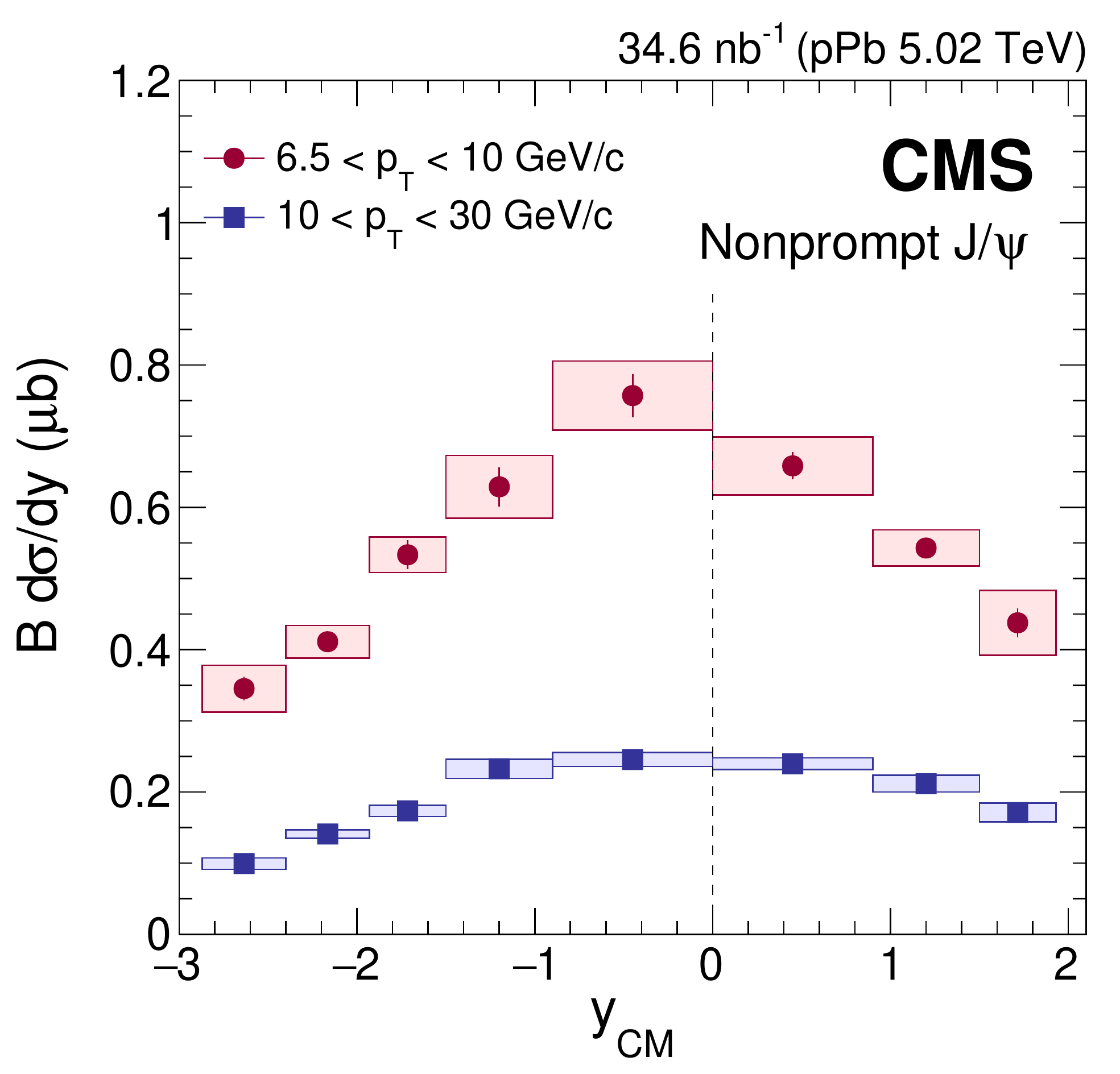}
\caption{Rapidity dependence of the cross section (multiplied by the dimuon branching fraction) for nonprompt \JPsi mesons in the \pt intervals of $6.5<\pt<10\GeVc$ (circles) and $10<\pt<30\GeVc$ (squares) in \pp (\cmsLeft) and \pPb (\cmsRight) collisions. The vertical dashed line indicates $\ycm=0$. The vertical bars (smaller than the symbols in most cases) represent the statistical uncertainties and the shaded boxes show the systematic uncertainties. The fully correlated global uncertainty from the integrated luminosity determination, 2.3\% for \pp and 3.5\% for \pPb collisions, is not included in the point-by-point uncertainties.}
\label{fig:cross_rap_np}
\end{figure}

\begin{figure*}[hbtp]
\centering
\includegraphics[width=0.98\textwidth]{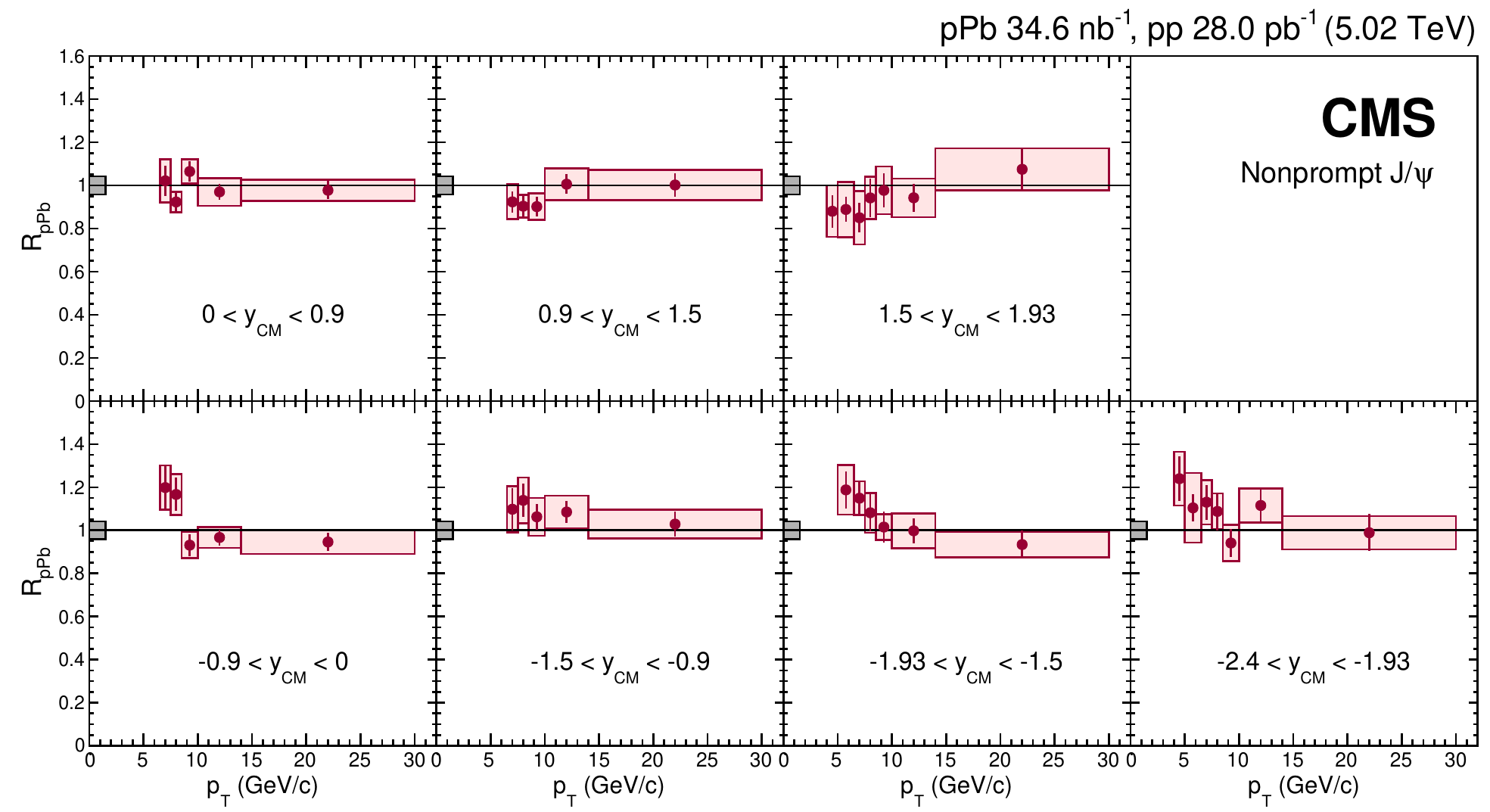}
\caption{Transverse momentum dependence of \rppb for nonprompt \JPsi mesons in seven \ycm ranges. The vertical bars represent the statistical uncertainties and the shaded boxes show the systematic uncertainties. The fully correlated global uncertainty of 4.2\% is displayed as a gray box at $\rppb=1$ next to the left axis.}
\label{fig:rppb_pt_np}
\end{figure*}

\begin{figure}[hbtp]
\centering
\includegraphics[width=0.48\textwidth]{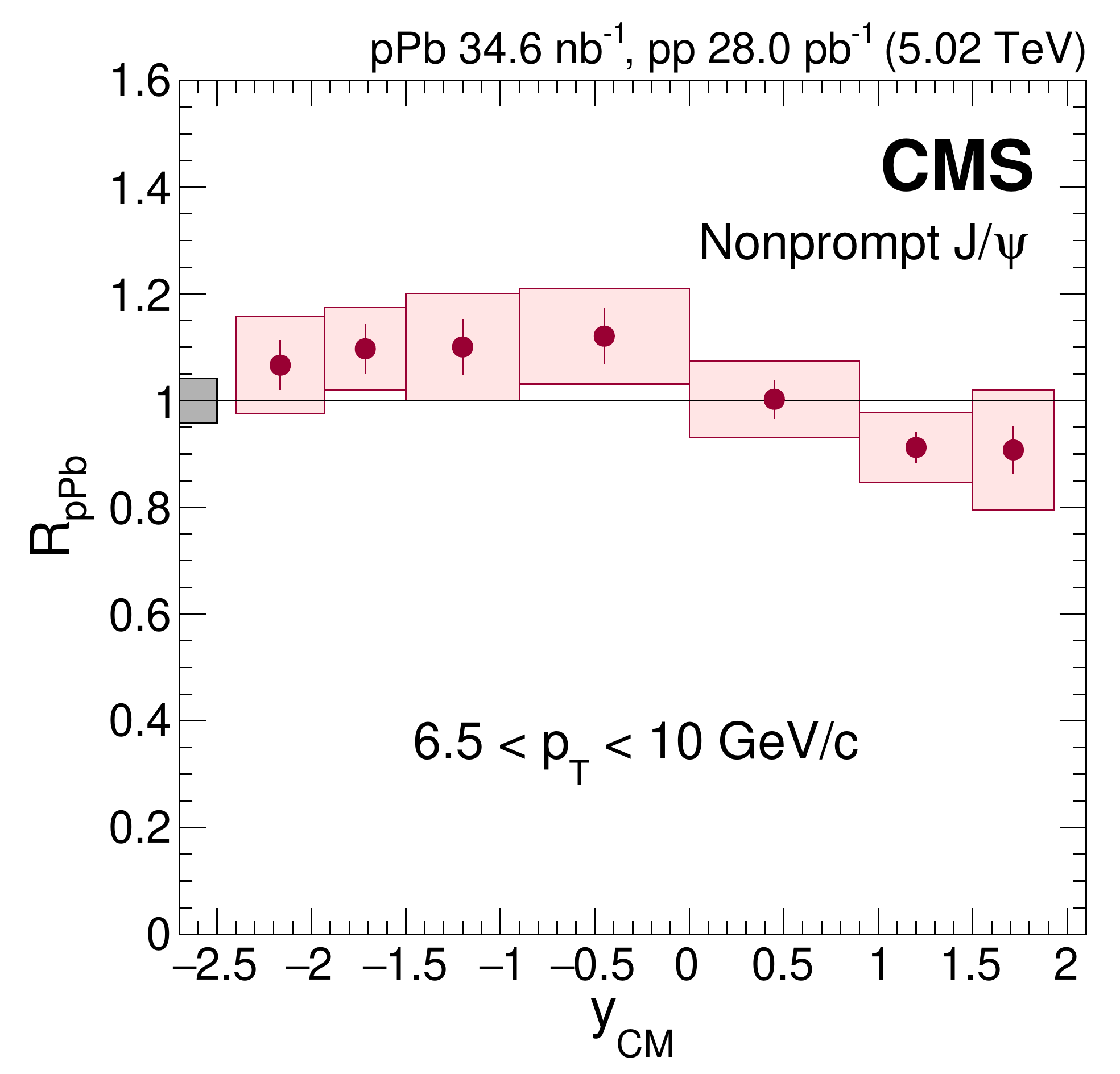}
\includegraphics[width=0.48\textwidth]{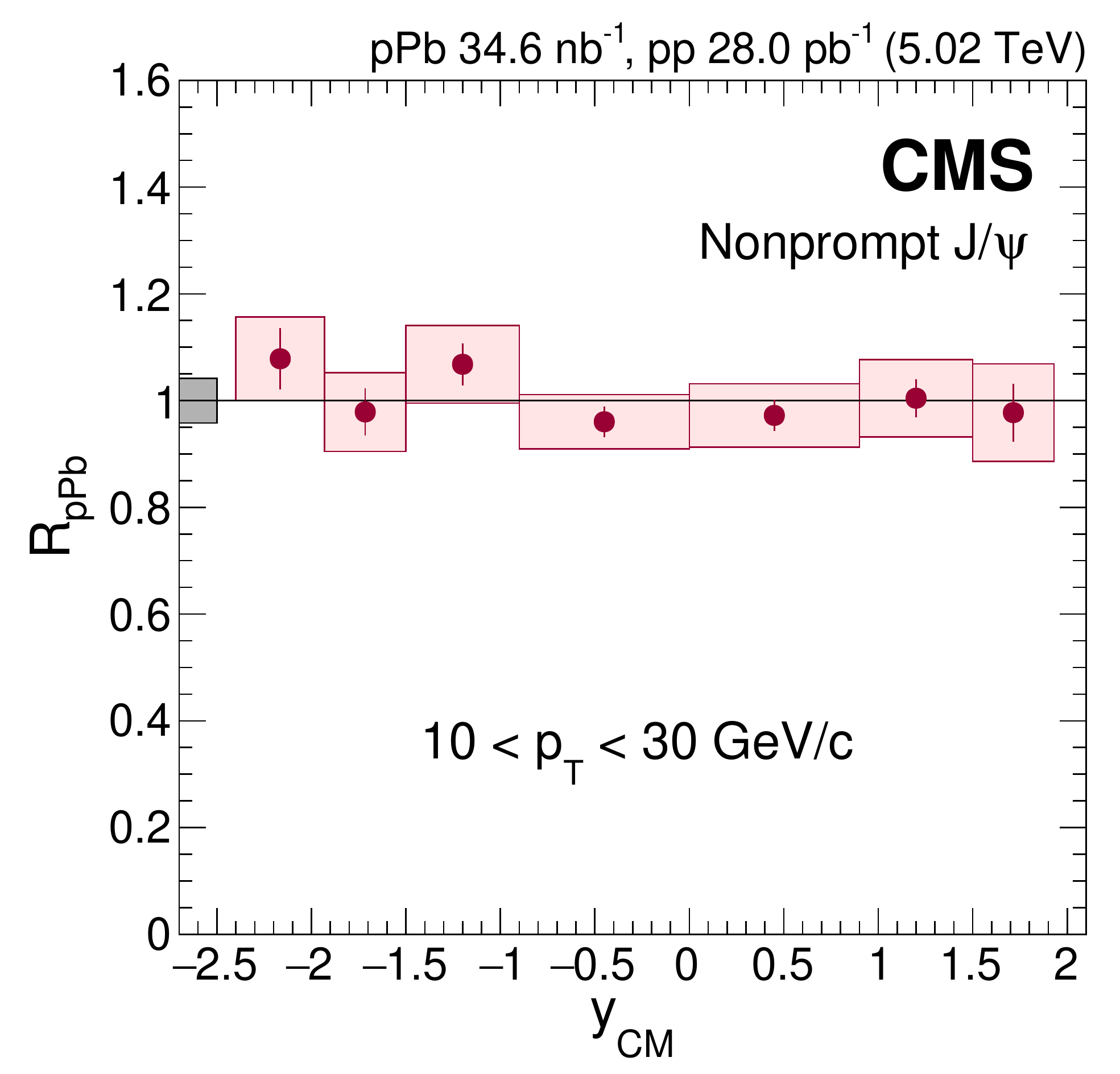}
\caption{Rapidity dependence of \rppb for nonprompt \JPsi mesons in two \pt ranges: $6.5<\pt<10\GeVc$ (\cmsLeft) and $10<\pt<30\GeVc$ (\cmsRight). The vertical bars represent the statistical uncertainties and the shaded boxes show the systematic uncertainties. The fully correlated global uncertainty of 4.2\% is displayed as a gray box at $\rppb=1$ next to the left axis.}
\label{fig:rppb_rap_np}
\end{figure}

\begin{figure}[hbtp]
\centering
\includegraphics[width=0.32\textwidth]{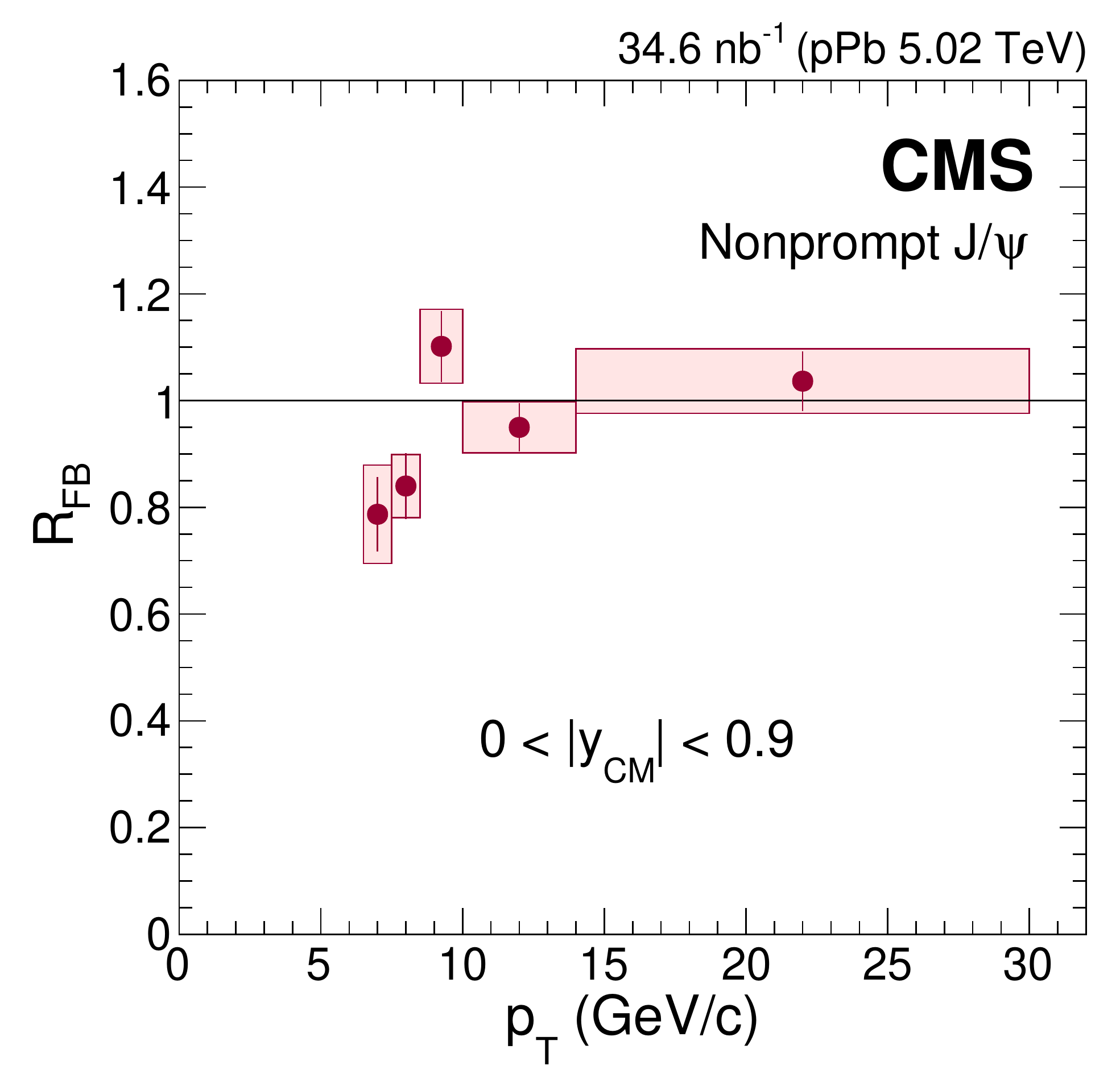}
\includegraphics[width=0.32\textwidth]{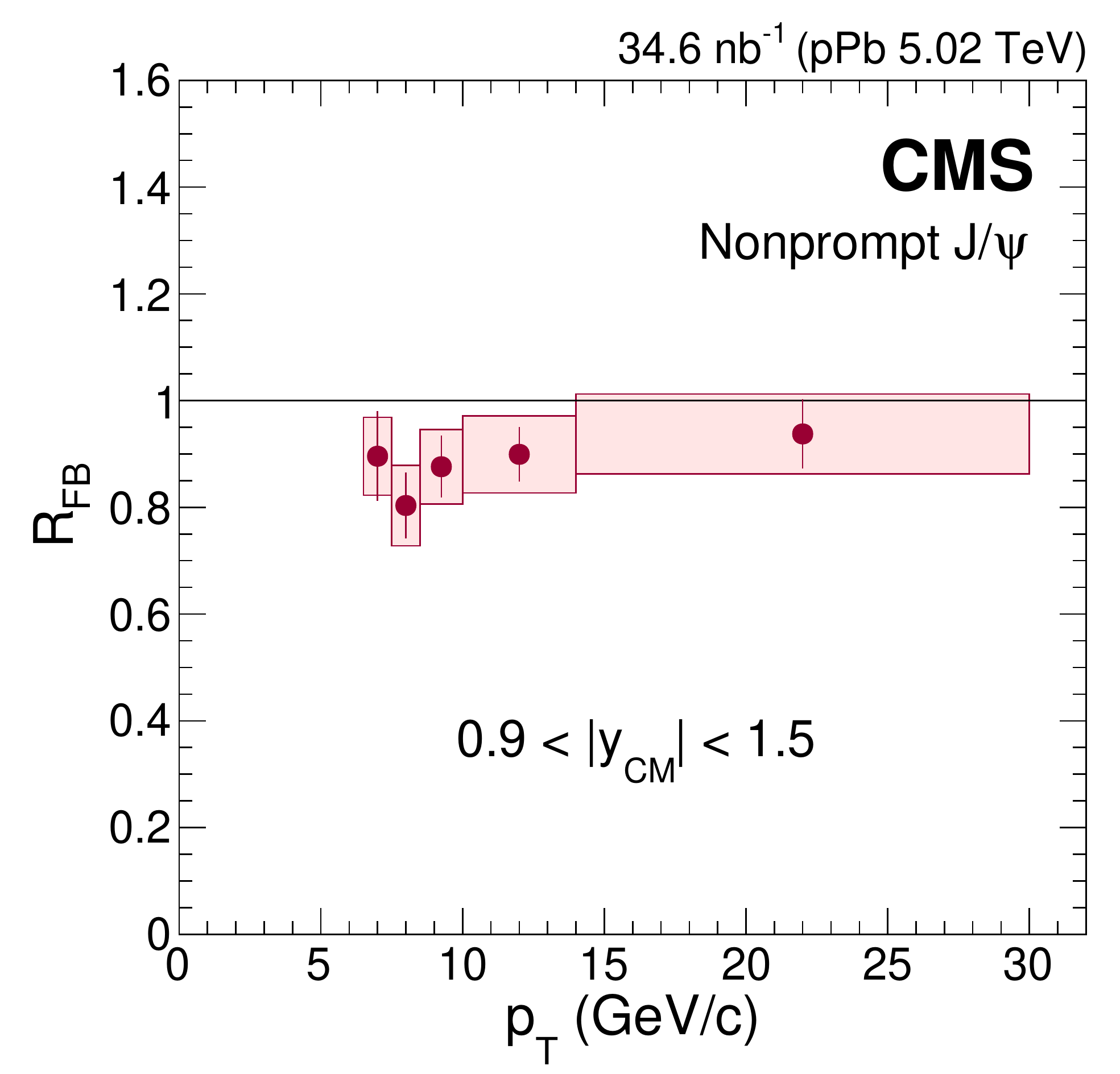}
\includegraphics[width=0.32\textwidth]{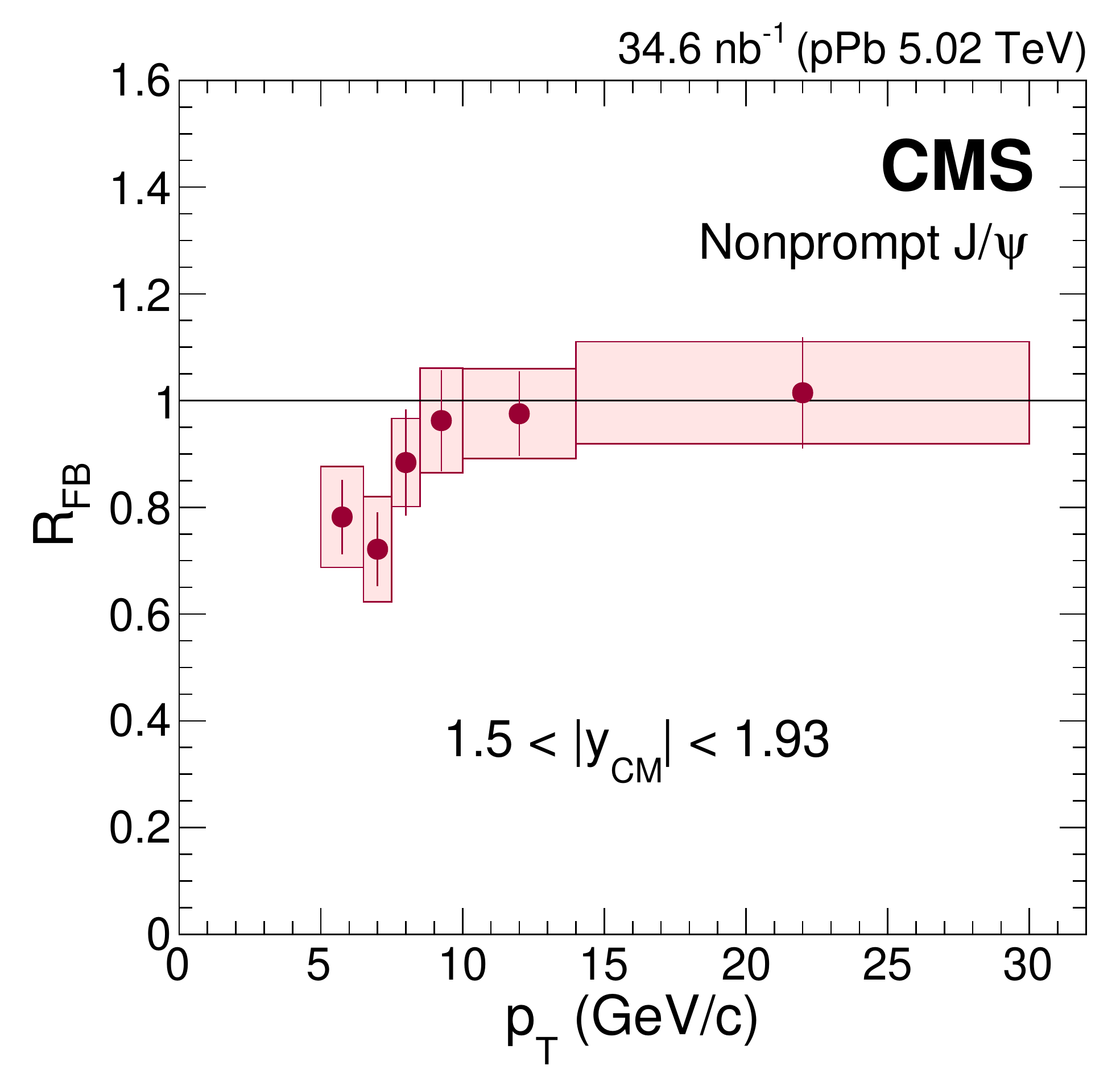}
\caption{Transverse momentum dependence of \rfb for nonprompt \JPsi mesons in three \ycm regions. The vertical bars represent the statistical uncertainties and the shaded boxes show the systematic uncertainties.}
\label{fig:rfb_pt_np}
\end{figure}

\begin{figure}[hbtp]
\centering
\includegraphics[width=0.48\textwidth]{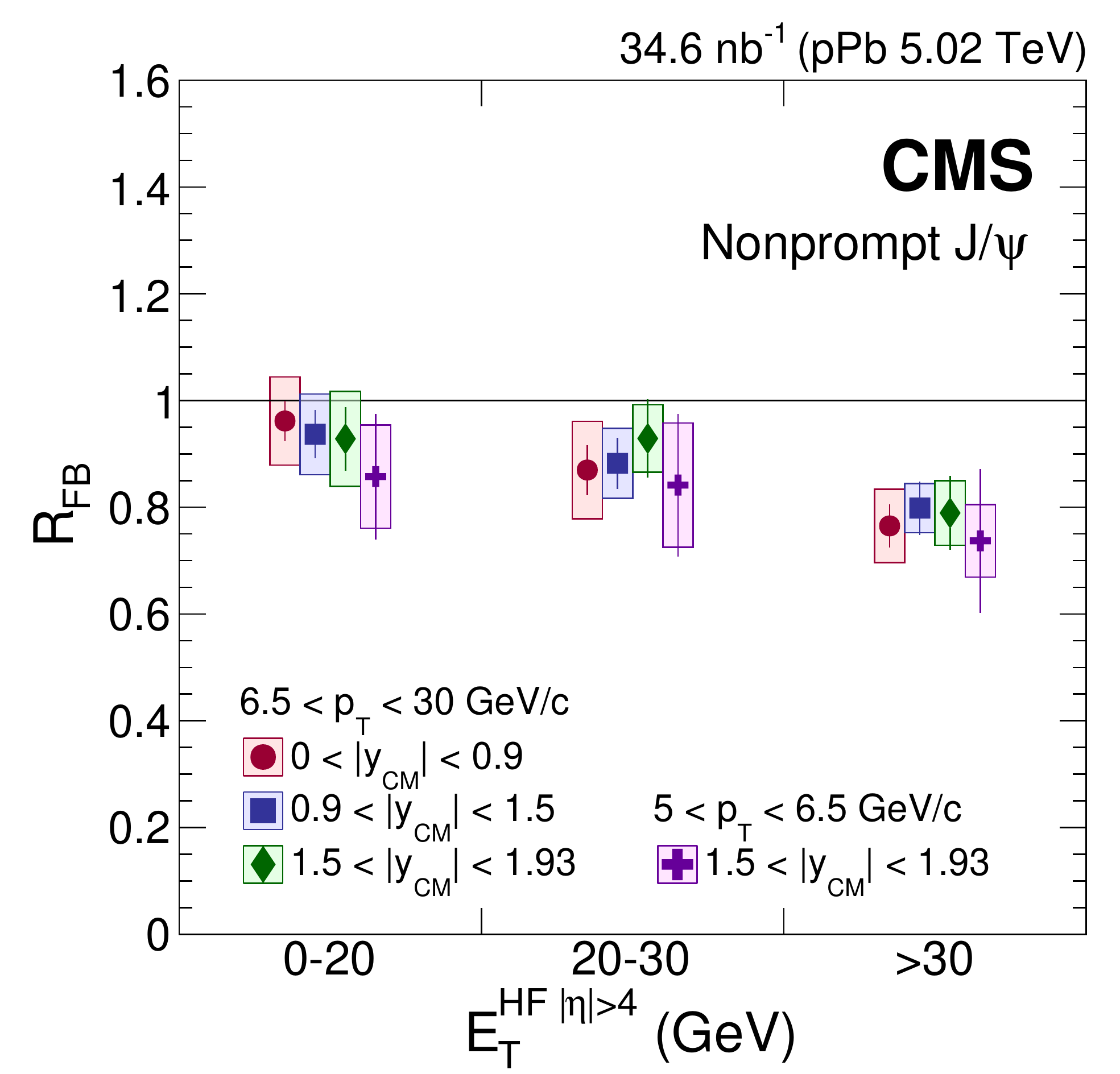}
\caption{Dependence of \rfb for nonprompt \JPsi mesons on the hadronic activity in the event, given by the transverse energy deposited in the CMS detector at large pseudorapidities \ethf. Data points are slightly shifted horizontally so that they do not overlap. The vertical bars represent the statistical uncertainties and the shaded boxes show the systematic uncertainties.}
\label{fig:rfb_ethf_np}
\end{figure}

{\tolerance=1200
The same distributions and observables discussed in Section~\ref{sec:promptjpsi} have been investigated for the nonprompt \JPsi meson samples. Differential cross sections are plotted as functions of \pt and \ycm in Figs.~\ref{fig:cross_pt_np} and~\ref{fig:cross_rap_np}, respectively, using the same binning as for prompt \JPsi mesons.
\par}

The measurement of \rppb for nonprompt \JPsi mesons shown in Fig.~\ref{fig:rppb_pt_np} as a function of \pt is compatible with unity in all \ycm bins. The somewhat larger uncertainties, however, make it difficult to draw firm conclusions for the nonprompt \JPsi production. The \ycm dependence of nonprompt \JPsi \rppb integrated in the low- and high-\pt regions is shown in Fig.~\ref{fig:rppb_rap_np}. In all \ycm bins, \rppb is consistent with unity although the data hint at a rapidity dependence for \rppb in the low \pt region, as found in the prompt \JPsi meson production (Fig.~\ref{fig:rppb_rap_pr}).

Figures~\ref{fig:rfb_pt_np} and~\ref{fig:rfb_ethf_np} show the \pt and \ethf dependence of nonprompt \JPsi \rfb, respectively. The \rfb ratios seem to increase slightly with \pt from ${\sim}0.8\pm0.1$ to ${\sim}1.0\pm0.1$ in all \ycm bins. The results are consistent with those from the ATLAS~\cite{Aad:2015ddl} and LHCb~\cite{Aaij:2013zxa} collaborations within uncertainties. As seen for prompt \JPsi meson production, \rfb for nonprompt \JPsi meson production decreases with \ethf, indicating the presence of different nuclear effects at forward than at backward \ycm in the regions with the greatest event activity.

\section{Summary}
\label{sec:summary}

Proton-proton (\pp) and proton-lead (\pPb) data at $\sqrtsnn=5.02\TeV$ collected with the CMS detector are used to investigate the production of prompt and nonprompt \JPsi mesons and its possible modification due to cold nuclear matter effects. Double-differential cross sections, as well as the nuclear modification factor \rppb and forward-to-backward production ratio \rfb, are reported as functions of the \JPsi \pt and \ycm.

The \rppb values for prompt \JPsi mesons are above unity in mid- and backward \ycm intervals analyzed ($-2.4<\ycm<0.9$), with a possible depletion in the most forward bin at low $\pt\lesssim7.5\GeVc$. In the case of nonprompt \JPsi meson production, \rppb is compatible with unity in all \ycm bins. The prompt \JPsi \rfb is below unity for $\pt\lesssim7.5\GeVc$ and forward $\abs{\ycm}>0.9$, but is consistent with unity for $\pt\gtrsim10\GeVc$. For nonprompt \JPsi mesons, \rfb tends to be below unity at $\pt\lesssim7.5\GeVc$ and increases for higher \pt, but with slightly larger uncertainties. The dependence of \rfb on the hadronic activity in \pPb events has been studied through the variable \ethf, characterizing the transverse energy deposited in the CMS detector at large pseudorapidities $4<\abs{\eta}<5.2$. The \rfb ratio is observed to decrease with increasing event activity for both prompt and nonprompt \JPsi mesons, indicating enhanced nuclear matter effects for increasingly central \pPb collisions.

A depletion of prompt \JPsi mesons in \pPb collisions (as compared to \pp collisions) is expected in the forward \ycm region because of the shadowing of nuclear parton distributions and/or coherent energy loss effects. Such a suppression is observed in the measurements presented in this paper at $\ycm>1.5$ and $\pt\lesssim7.5\GeVc$, but not at larger \pt, consistent with the expected reduced impact of nuclear parton distributions and coherent energy loss effects for increasing \JPsi \pt. At negative \ycm, both shadowing and energy loss effects are known to lead to small nuclear modifications, as confirmed by the present measurements. Such processes are also expected to affect the nuclear dependence of $\PB$ hadron production and thereby, through its decays, nonprompt \JPsi production. The measurements presented here provide new constraints on cold nuclear matter effects on prompt and nonprompt \JPsi production over a wide kinematic range.

\begin{acknowledgments}
We congratulate our colleagues in the CERN accelerator departments for the excellent performance of the LHC and thank the technical and administrative staffs at CERN and at other CMS institutes for their contributions to the success of the CMS effort. In addition, we gratefully acknowledge the computing centers and personnel of the Worldwide LHC Computing Grid for delivering so effectively the computing infrastructure essential to our analyses. Finally, we acknowledge the enduring support for the construction and operation of the LHC and the CMS detector provided by the following funding agencies: BMWFW and FWF (Austria); FNRS and FWO (Belgium); CNPq, CAPES, FAPERJ, and FAPESP (Brazil); MES (Bulgaria); CERN; CAS, MoST, and NSFC (China); COLCIENCIAS (Colombia); MSES and CSF (Croatia); RPF (Cyprus); SENESCYT (Ecuador); MoER, ERC IUT, and ERDF (Estonia); Academy of Finland, MEC, and HIP (Finland); CEA and CNRS/IN2P3 (France); BMBF, DFG, and HGF (Germany); GSRT (Greece); OTKA and NIH (Hungary); DAE and DST (India); IPM (Iran); SFI (Ireland); INFN (Italy); MSIP and NRF (Republic of Korea); LAS (Lithuania); MOE and UM (Malaysia); BUAP, CINVESTAV, CONACYT, LNS, SEP, and UASLP-FAI (Mexico); MBIE (New Zealand); PAEC (Pakistan); MSHE and NSC (Poland); FCT (Portugal); JINR (Dubna); MON, RosAtom, RAS, RFBR and RAEP (Russia); MESTD (Serbia); SEIDI, CPAN, PCTI and FEDER (Spain); Swiss Funding Agencies (Switzerland); MST (Taipei); ThEPCenter, IPST, STAR, and NSTDA (Thailand); TUBITAK and TAEK (Turkey); NASU and SFFR (Ukraine); STFC (United Kingdom); DOE and NSF (USA).

\hyphenation{Rachada-pisek} Individuals have received support from the Marie-Curie program and the European Research Council and EPLANET (European Union); the Leventis Foundation; the A. P. Sloan Foundation; the Alexander von Humboldt Foundation; the Belgian Federal Science Policy Office; the Fonds pour la Formation \`a la Recherche dans l'Industrie et dans l'Agriculture (FRIA-Belgium); the Agentschap voor Innovatie door Wetenschap en Technologie (IWT-Belgium); the Ministry of Education, Youth and Sports (MEYS) of the Czech Republic; the Council of Science and Industrial Research, India; the HOMING PLUS program of the Foundation for Polish Science, cofinanced from European Union, Regional Development Fund, the Mobility Plus program of the Ministry of Science and Higher Education, the National Science Center (Poland), contracts Harmonia 2014/14/M/ST2/00428, Opus 2014/13/B/ST2/02543, 2014/15/B/ST2/03998, and 2015/19/B/ST2/02861, Sonata-bis 2012/07/E/ST2/01406; the National Priorities Research Program by Qatar National Research Fund; the Programa Clar\'in-COFUND del Principado de Asturias; the Thalis and Aristeia programs cofinanced by EU-ESF and the Greek NSRF; the Rachadapisek Sompot Fund for Postdoctoral Fellowship, Chulalongkorn University and the Chulalongkorn Academic into Its 2nd Century Project Advancement Project (Thailand); and the Welch Foundation, contract C-1845. \end{acknowledgments}

\clearpage
\bibliography{auto_generated}
\cleardoublepage \appendix\section{The CMS Collaboration \label{app:collab}}\begin{sloppypar}\hyphenpenalty=5000\widowpenalty=500\clubpenalty=5000\input{HIN-14-009-authorlist.tex}\end{sloppypar}
\end{document}

%% file: HIN-14-009-authorlist.tex
\textbf{Yerevan Physics Institute,  Yerevan,  Armenia}\\*[0pt]
A.M.~Sirunyan, A.~Tumasyan
\vskip\cmsinstskip
\textbf{Institut f\"{u}r Hochenergiephysik,  Wien,  Austria}\\*[0pt]
W.~Adam, E.~Asilar, T.~Bergauer, J.~Brandstetter, E.~Brondolin, M.~Dragicevic, J.~Er\"{o}, M.~Flechl, M.~Friedl, R.~Fr\"{u}hwirth\cmsAuthorMark{1}, V.M.~Ghete, C.~Hartl, N.~H\"{o}rmann, J.~Hrubec, M.~Jeitler\cmsAuthorMark{1}, A.~K\"{o}nig, I.~Kr\"{a}tschmer, D.~Liko, T.~Matsushita, I.~Mikulec, D.~Rabady, N.~Rad, B.~Rahbaran, H.~Rohringer, J.~Schieck\cmsAuthorMark{1}, J.~Strauss, W.~Waltenberger, C.-E.~Wulz\cmsAuthorMark{1}
% ...
\vskip\cmsinstskip
% ...
1:~~Also at Vienna University of Technology, Vienna, Austria\\
% ...
